\documentclass{article}
\usepackage[utf8]{inputenc}

\usepackage{lmodern}

\usepackage{amsmath, bm}        % extensions for typesetting of math
\usepackage{amsfonts}       % math fonts
\usepackage{amsthm}         % theorems, definitions, etc.
\usepackage{bbding}         % various symbols (squares, asterisks, scissors, ...)
\usepackage{bm}             % boldface symbols (\bm)
\usepackage{graphicx}       % embedding of pictures
\usepackage{fancyvrb}       % improved verbatim environment
\usepackage{bbm} %%19.8.25 e.g. $\mathbbm{R}$ 

\usepackage{dsfont}
\usepackage[dvipsnames]{xcolor}
\usepackage{csquotes}
\usepackage{bigints}
\usepackage{appendix}
\usepackage{tipa}
\usepackage{tensor}
\usepackage{hhline}
\usepackage{multirow}
\usepackage{diagbox}
\usepackage{makecell}
\usepackage{float}

\usepackage{geometry}
\def \bl {\mbox{\boldmath{$\ell$}}}		
\def \bn {\mbox{\boldmath{$n$}}}

\newcommand{\beq}{\begin{eqnarray}}
\newcommand{\eeq}{\end{eqnarray}}
\newcommand{\be}{\begin{equation}}
\newcommand{\ee}{\end{equation}}
\newcommand{\beqn}{\begin{eqnarray}}
\newcommand{\eeqn}{\end{eqnarray}}
\newcommand{\pa}{\partial}
\newcommand{\OO}[1] {{O}(r^{-#1})}
\newcommand{\OOp}[1] {{O}(r^{#1})}

\newcommand{\M}[3] {{\stackrel{#1}{M}}_{{#2}{#3}}}
\newcommand{\m}[3] {\!{\stackrel{\hspace{.3cm}#1}{m}}_{\!{#2}{#3}}\,}
\def\d{\mathrm{d}}
\def \bF {\mbox{\boldmath{$F$}}}
\def \bA {\mbox{\boldmath{$A$}}}

\usepackage{authblk} %for authors with multiple affiliations
\usepackage{cite,enumerate}

    \newgeometry{vmargin={15mm}, hmargin={12mm,17mm}}

\newcommand*\mypsi{\Psi^\prime}
\newcommand*\myomega{\Omega^\prime}
\newcommand*\myd{\mathrm{d}}

\title{On the uniqueness of the Kerr-(A)dS metric as a type II(D) solution \linebreak in six dimensions}

\author[1,2]{David Koko\v ska\thanks{david.kokoska@matfyz.cz}}
\author[2]{Marcello Ortaggio\thanks{ortaggio(at)math(dot)cas(dot)cz}}

\affil[1]{Institute of Theoretical Physics, Faculty of Mathematics and Physics, \newline
 Charles University, V Hole\v{s}ovi\v{c}k\'{a}ch 2, 180 00 Prague 8, Czech Republic}

\affil[2]{Institute of Mathematics, Czech Academy of Sciences, \newline \v Zitn\' a 25, 115 67 Prague 1, Czech Republic}

\begin{document}

\maketitle

\abstract{We study the class of six-dimensional $\Lambda$-vacuum spacetimes which admit a non-degenerate multiple Weyl aligned null direction $\bl$ (thus being of Weyl type~II or more special) with a ``generic'' optical matrix. Subject to an additional assumption on the asymptotic fall-off of the Weyl tensor, we obtain the most general metric of this class, which is specified by one discrete (normalized) and three continuous parameters. All solutions turn out to be Kerr--Schild spacetimes of type~D and, in passing, we comment on their Kerr--Schild double copy. We further show that the obtained family is locally isometric to the general doubly-spinning Kerr-NUT-(A)dS metric with the NUTs parameters switched off. 
In particular, the Kerr-(A)dS subclass and its extensions (i.e., analytic continuation and ``infinite-rotation'' limit) are recovered when certain polynomial metric functions are assumed to be fully factorized. As a side result, a unified metric form which encompasses all three branches of the extended Kerr-(A)dS family in all even dimensions is presented in an appendix.}

\maketitle

\tableofcontents

\section{Introduction and summary of results}

\subsection{Background}

\label{subsec_backgr}

The Kerr and Kerr-Newman metrics, as well as their generalizations to include a cosmological constant or non-spherical topologies, and their static limits (cf., e.g., \cite{Stephanibook,GriPodbook} and references therein) describe various black holes which share two remarkable properties. First, they are all of Petrov type~D. Secondly, they admit a Kerr--Schild(-AdS) representation. However, they exhaust neither the type~D \cite{Carter68pla,Carter68cmp,Kinnersley69,Debever71,Carter73,Plebanski75_2,PleDem76,GarciaD84,DebKamMcL84} nor the Kerr--Schild (KS) \cite{KerSch65,KerSch652,DebKerSch69} classes of solutions, for both of which all vacuum solutions are known.\footnote{To be precise, ref.~\cite{DebKerSch69} studied only the case of a vanishing cosmological constant $\Lambda$. Throughout the paper, for the sake of brevity, we shall call KS spacetimes both those without and with $\Lambda$.} 

While displaying several qualitatively new features, also the higher-dimensional extensions of such black holes \cite{Tangherlini63,MyePer86,Chakrabarti86,HawHunTay99,Gibbonsetal04} possess the same properties, being again KS (in fact, by construction \cite{MyePer86,Gibbonsetal04}) and of Weyl type~D \cite{Pravdaetal04,ColPel06,Hamamotoetal07,PraPraOrt07,OrtPraPra09,OrtPodZof08} (in the classification scheme of \cite{Coleyetal04}, cf. also the reviews \cite{Coley08,OrtPraPra13rev}).\footnote{A notable violation of both conditions occurs in the case of the five-dimensional black ring of \cite{EmpRea02prl}, as discussed in \cite{PraPra05,OrtPraPra09}.} Similarly as in the four dimensional case, it would thus be desirable to clarify what is the role played by the black holes of \cite{Tangherlini63,MyePer86,Chakrabarti86,HawHunTay99,Gibbonsetal04} within both the KS and the type~D (or, more generally, type~II) classes of vacuum solutions, and to what extent they possibly exhaust those.\footnote{Over the past decade, renewed interest in KS spacetimes has been also motivated by the KS formulation of the classical double copy \cite{MonOCoWhi14} -- cf. section~\ref{subsubsec_DC} for some comments and references.}
Various general results about $n$-dimensional KS spacetimes have been obtained in \cite{DerGur86,ColHilSen01,OrtPraPra09,MalPra11,OrtSri24,Srinivasan25}, while several properties of type~II/D metrics in higher dimensions have been elucidated in \cite{PraPraOrt07,Durkee09,DurRea09,ParWyl11,Ortaggioetal12,OrtPraPra13,ReaGraTur13,deFGodRea15,Wylleman15,deFGodRea16,Ortaggio17,OrtPraPra18,TinPra19,Taghavi-Chabert22}. In particular, all five-dimensional vacua (including a cosmological constant) of type~II have been fully classified in \cite{deFGodRea15,Wylleman15,deFGodRea16} (see also \cite{ParWyl11,ReaGraTur13} for earlier results in special cases) according to the possible rank of the optical matrix (defined below in \eqref{L_def}) associated to a multiple Weyl aligned null direction (mWAND). Remarkably, in the {\em full-rank} case \cite{deFGodRea15}, such solutions reduce to the 5D version of Kerr-(A)dS black hole \cite{HawHunTay99} (including an analytic continuation and a limit thereof), thereby being of type~D (thus ruling out the type~II in this class) and specified just by a few parameters -- this contrasts with the more numerous solutions in four dimensions \cite{Stephanibook,GriPodbook}. Furthermore, the class of metrics obtained in \cite{deFGodRea15} subsequently turned out (for $\Lambda>0$) to be equivalent to the ``Kerr-de~Sitter-like'' class (a subset of KS metrics) \cite{MarPeo22}. 

It would clearly be interesting to extend the results of \cite{deFGodRea15,Wylleman15,deFGodRea16} to arbitrary dimensions. Not surprisingly, however, this task appears to be considerably more complicated. First, the analysis \cite{deFGodRea15,Wylleman15,deFGodRea16} relied on a five-dimensional extension of the Goldberg-Sachs theorem \cite{Pravdaetal04,DurRea09,Ortaggioetal12}, which is not yet available in arbitrary dimension (see, however, \cite{Pravdaetal04,OrtPraPra09,DurRea09,OrtPraPra09b,OrtPraPra13}, and major progress in six dimensions in \cite{TinPra19}). Secondly, a qualitative difference in the structure of the Weyl tensor appears when moving from $n=4,5$ to $n=6$ (or any higher dimensions), amounting to new ``degrees of freedom'' in the purely spatial components $C_{ijkl}$ \cite{PraPraOrt07,OrtPra14}. This makes the integration of the Newman-Penrose (NP) equations more involved, and may require new techniques. Nevertheless, based on earlier results in various special cases \cite{Pravdaetal04,PodOrt06,OrtPraPra09b,OrtPraPra13}, one might expect that a ``uniqueness'' result similar to that of \cite{deFGodRea15} may hold also in more than five dimensions, when the optical matrix has full-rank.
Indeed, this has been proven in six dimensions \cite{Ortaggio17}, but only when $\Lambda=0$, and under an additional assumption (cf.~\eqref{bw0WeylAsymBehaviour} below) on the asymptotic fall off of the Weyl tensor (which is identically satisfied for $n=4,5$). The main purpose of the present contribution is thus to extend the classification of six-dimensional type~II/D vacua performed in \cite{Ortaggio17} to allow for a non-zero cosmological constant, and hopefully pave the way for a complete extension to any dimension. Apart from making the analysis of the NP equations considerably more complicated, as well as changing the global structure of the spacetime, the presence of a cosmological constant brings in additional new features such as a richer structure of the roots of certain metric functions (defining, e.g., axes of symmetry or Killing horizons), which is in turn related to the existence of specific ``ultraspinning'' \cite{HawHunTay99,CalEmpRod08,ArmObe11,Caldarellietal12,Gnecchietal14,Klemm14,HenManKub15,Hennigaretal15} or ``infinite-rotation'' configurations \cite{MarPaeSen17,MarPaeSen18,MarPeo22_b}. These limiting configurations will be naturally comprised by our analysis, and as a byproduct we will thus also obtain a unified form of various Einstein spacetimes previously presented using different coordinates and parametrizations, with the advantage that no limits or analytic continuations will now be needed to connect them.  
 
In the remaining part of this section we describe all the assumptions made throughout our contribution, summarize our findings, and outline the structure of the paper.

\subsection{Summary of results}

\label{subsec_summary}

In this work we classify all the six-dimensional Einstein spacetimes which obey the following assumptions (some more technical details will be given in section~\ref{subsec_assumpt}):

\begin{enumerate}

\item\label{ass1} The Weyl tensor is of type~II or more special in the classification of \cite{Coleyetal04}, i.e. \cite{Ortaggio09},
\be
	\ell_{[e}C_{a]b[cd}\ell_{f]}\ell^b=0 ,
	\label{type_II}
\ee
where $\bl$ is a null vector field.

\item\label{ass2} The corresponding mWAND $\bl$ (which can be assumed to be geodesic w.l.g. \cite{DurRea09}) possesses a {\em non-degenerate} optical matrix (eq.~\eqref{DetofL}). 

\item\label{ass2b} In addition to being non-degenerate, the optical matrix is ``generic'', i.e., the (modules of the) eigenvalues of its antisymmetric part (cf.~\eqref{Lin6D}) are distinct and both non-constant (this assumption will only be enforced from section~\ref{sec_generic} on).

\item\label{ass3} The spatial part of the Weyl tensor falls off ``fast enough'' far away along an affine parameter of $\bl$ (eq.~\eqref{bw0WeylAsymBehaviour}).

\end{enumerate}

Assumption~\ref{ass1}. has been already put in due context in section~\ref{subsec_backgr}. Assumption~\ref{ass2}. is typically relevant to black hole solutions and to asymptotically simple spacetimes (as opposed, e.g., to black strings, which correspond to a degenerate optical matrix -- cf. \cite{OrtPraPra09,OrtPraPra09b,OrtPraPra11,ReaGraTur13} for related comments). The genericity assumption~\ref{ass2b}. has mainly a technical character. It enables us to exclude from the present analysis certain special (``non-generic'') subcases which require a separate lengthy analysis and will deserve an independent investigation elsewhere.\footnote{Based on previous works in related contexts \cite{deFGodRea15,Ortaggio17}, one might expect the non-generic subcases to include, for example, metrics with a single spin or with two equal spins, as opposed to the generic case with two unequal spins analyzed in the present contribution.}
The motivation for assumption~\ref{ass3}. is twofold. From a physical viewpoint, for $\lambda<0$ such a fall off is required to describe $n>4$ asymptotically AdS spacetimes (for which it implies the conformal boundary is conformally flat) \cite{AshDas00}. Additionally, in any dimension it allows for a partial extension of the Goldberg-Sachs theorem \cite{OrtPraPra09b}, technically resulting in eq.~\eqref{NewOpticalMatrixInverse}, which in turn leads to a significant simplification of the NP equations and ultimately enables a complete integration.\footnote{Mathematically speaking, such an assumption means that certain integration functions (when fixing the Weyl $r$-dependence from the Bianchi identities) are set to be zero \cite{OrtPra14} (cf. also, e.g., section~5 of \cite{PraPra08} and section~4 of \cite{OrtPraPra13}). It should be pointed out, however, that a drawback of this simplification is that it leads to disregarding solutions such as static black holes with a generic Einstein horizon \cite{GibWil87,Birmingham99} or rotating black holes with non-zero NUT \cite{CheLuPop06} (as follows from \cite{PodOrt06} and \cite{Hamamotoetal07}, respectively). On the other hand, assumption~\ref{ass3}. is automatically satisfied by Kerr--Schild spacetimes \cite{OrtPraPra09,MalPra11}.}

Under the assumptions listed above, our results can be concisely described as follows. 

\begin{enumerate}[a.]

\item We obtain the explicit, local form of all the corresponding metrics, given by \eqref{CarterPlebanskiEF} with \eqref{Q_main}. These can be fully classified in terms of four continuous parameters (only three being essential thanks to a scaling freedom) and, as we further show, turn out to be locally isometric to the zero-NUT subclass of the Einstein spacetimes constructed in \cite{CheLuPop06}. A special subfamily is singled out by assuming the metric function ${\cal P}(s)$ to be fully factorisable, resulting in the generalized Kerr-(A)dS solution with two unequal spins~\eqref{KerrAdS} (equivalent to~\eqref{FinalMetricForU0neq0andNonEqualRotations}).
In this case, depending on the value of
a discrete parameter $\epsilon=\pm1, 0$ (cf. the comments around~\eqref{epsilon_Definition}) one recovers the Kerr-(A)dS metric \cite{Gibbonsetal04} and its generalizations \cite{MarPeo22_b,ChrConGra25}, now expressed in a unified way.
 The various line elements are obtained in Eddington--Finkelstein-like coordinates $(u,r,x^\alpha)$, which are natural in the NP formalism employed in the present paper. However, in appendix~\ref{Section: Appendix: Kerr-(A)dS metrics in six dimensions} we present various alternative coordinate systems more frequently used in the black hole literature.

\item All metrics of the considered class are in fact of Weyl type~D (i.e., the genuine type~II turns out to be forbidden under the assumptions).

\item All metrics of the considered class turn out to admit a KS form. As a side result, this implies that all six-dimensional KS Einstein spacetimes admitting a non-degenerate KS and subject to assumption~\ref{ass2b}. are thus now known, extending the Ricci-flat result of \cite{Ortaggio17}  (recall \cite{OrtPraPra09,MalPra11} that the KS vector of KS Einstein spacetimes is necessarily a geodesic mWAND, and condition~\ref{ass3}. is also met by such spacetimes). Cf.~\cite{deFGodRea15,MarPeo22} in five dimensions.

\end{enumerate}

The plan of the paper is as follows. In section~\ref{sec_r_dep} (under assumptions~\ref{ass1}., \ref{ass2}. and \ref{ass3}. listed above), we obtain the $r$-dependence of the non-zero NP quantities, along with the ``transverse'' equations to be employed subsequently. In section~\ref{sec_adapted} we define adapted coordinates and a natural parallelly transported frame (the results of sections~\ref{sec_r_dep} and \ref{subsec_coord_u} hold in arbitrary dimension $n>4$, while from section~\ref{subsec_mi} on we specialize to $n=6$). Under the additional assumption~\ref{ass2b}., the complete integration of the field equations is carried out in section~\ref{sec_generic}: the most general metric is obtained in subsection~\ref{subsec_general_sol}, while subsection~\ref{subsubsec_recovering} is devoted to identifying the special subclass corresponding to the doubly spinning Kerr-(A)dS black holes of \cite{Gibbonsetal04} and their extensions \cite{MarPeo22_b,ChrConGra25}. Appendix~\ref{app_NP} summarizes concisely the higher dimensional NP formalism used in the paper. Appendix~\ref{app_Phi_6D} contains some technicalities to be used in the main text, namely we determine the exact $r$-dependence of the Weyl components $\Phi_{ij}$, which in turn (cf.~\eqref{U}) leads to the exact $r$-dependence of the line-element~\eqref{InverseMetricin6D_General}. In appendix~\ref{Section: Appendix: Kerr-(A)dS metrics in six dimensions} we review the Kerr-(A)dS metric of \cite{Gibbonsetal04} in arbitrary even dimension and its generalizations \cite{MarPeo22_b,ChrConGra25}, presenting it in various coordinate systems similar to those of \cite{MyePer86}. For the particular case of $n=6$ doubly-spinning metrics, we additionally overview the alternative (``intrinsic'') spatial coordinates proposed in \cite{CheLuPop06}.

\paragraph{Notation}

Throughout the paper, we follow the by now standard higher-dimensional NP formalism \cite{Pravdaetal04,Coleyetal04vsi,OrtPraPra07,Durkeeetal10,OrtPraPra13rev}, cf. appendix~\ref{app_NP}. In particular, the non-zero Weyl components relevant to this paper are defined in~\eqref{Weyl0} and \eqref{Weyl<0}. Any object with a superscript $0$ will denote a quantity which is independent of the coordinate $r$. Indices run as follows (at a later stage we will set $n=6$):
\be
 a, b,c,\ldots=0,\ldots,n-1 , \qquad i, j,k,  \ldots=2,\ldots,n-1 . 
 \label{indices}
\ee

\section{Fixing the $r$-dependence, and transverse equation (any dimension)}

\label{sec_r_dep}

\subsection{Preliminaries: assumptions and choice of null frame}

\label{subsec_assumpt}

We will study $n$-dimensional ($n>4$) Einstein spacetimes\footnote{The constant $\lambda$ is just a convenient rescaling of the standard cosmological constant $\Lambda$, i.e., $\lambda=\frac{2\Lambda}{(n-1)(n-2)}$.}  
\be
	R_{ab}=(n-1)\lambda g_{ab} , \qquad \lambda \equiv \dfrac{R}{n(n-1)} ,
\ee
assumed to be of Weyl type II or more special in the classification of \cite{Coleyetal04} (see also \cite{OrtPraPra13rev}), i.e., possessing an mWAND $\bl$ (eq.~\eqref{type_II}).

Let us introduce a frame $\mbox{\boldmath{$m$}}_{(a)}$ adapted to $\bl$, i.e., consisting of two null vectors $\bl\equiv\mbox{\boldmath{$m$}}_{(0)}$,  $\bn\equiv\mbox{\boldmath{$m$}}_{(1)}$ and $n-2$ orthonormal spacelike vectors $\mbox{\boldmath{$m$}}_{(i)}$ (with~\eqref{indices}) \cite{Coleyetal04,OrtPraPra13rev}. 
In a spacetime admitting an mWAND, there always exists a {\em geodesic} mWAND \cite{DurRea09}. With no loss of generality we can thus assume $\bl$ to be geodesic and affinely parametrized, i.e., 
\begin{align}\label{GeodandAffineCondition}
    L_{i0} = 0 = L_{10}.
\end{align}

The \textit{optical matrix} of $\bm{\ell}$ is defined as \cite{Pravdaetal04,OrtPraPra07}
\begin{align}
    L_{ij} \equiv \ell_{a;b}m^{a}_{(i)}m^{b}_{(j)}.
		\label{L_def}
\end{align}
Throughout the paper we will restrict ourselves to the case of a {\em non-degenerate} optical matrix, i.e.,
\begin{align}\label{DetofL}
    \det L \neq 0.
\end{align}
This implies, in particular, that the Weyl type can only be II or D (or O, in the trivial case of (A)dS space) \cite{Pravdaetal04,Kubicek_thesis,OrtPraPra18}.

We can further choose the full null frame $(\bm{\ell}, \bm{n}, \bm{m}_{(i)})$  to be parallelly transported along $\bm{\ell}$ \cite{OrtPraPra07}, i.e.,
\begin{align}\label{ParaFrame}
    \overset{i}{M}_{j0} = 0 = N_{i0}.
\end{align}
Next, thanks to~\eqref{DetofL} we can use a null rotation to uniquely fix the null vector $\bm{n}$ such that \cite{OrtPraPra09} (see also \cite{deFGodRea15,Ortaggio17})
\begin{align}\label{Li1}
    L_{i1} = 0,
\end{align}
without affecting any of the previous conditions.

The Sachs equation \cite{Pravdaetal04,OrtPraPra07} for an Einstein spacetime with \eqref{GeodandAffineCondition}, \eqref{ParaFrame} reduces to
\begin{align}
    DL = -L^2 .
\end{align}
Thanks to \eqref{DetofL}, its solution reads \cite{OrtPraPra09,OrtPraPra09b}
\begin{align}
\label{OpticalMatrixInverse}
    L^{-1} = r\mathbbm{1} - b,
\end{align}
where $\mathbbm{1}$ is the $(n-2)\times(n-2)$ identity matrix, $b$ a matrix such that $Db=0$, and $r$ is an affine parameter along the geodesics of $\bm{\ell}$. Taking $r$ as one of the spacetime coordinates, it means 
\be
 \bm{\ell} = \partial_r .
\ee

Our final assumption concerns the fall-off behaviour of the spatial part of the Weyl tensor for large~$r$ (cf. section~\ref{subsec_summary} for related comments), which we require to be ``fast enough'', namely 
\begin{align}\label{bw0WeylAsymBehaviour}
    C_{ijkm} = o(r^{-2}) .
\end{align}

Thanks to~\eqref{bw0WeylAsymBehaviour}, using the Bianchi identities~\eqref{bw0DiffBianchiPhi}, \eqref{bw0DiffBianchiC} it follows \cite{OrtPraPra09b}
\begin{align}\label{Sym_b}
    b_{(ij)} = \dfrac{\mathrm{Tr}b}{n-2}\delta_{ij},
\end{align}
where $\mathrm{Tr}b \equiv b_{ii}$ is the trace of $b_{ij}$. The latter (with \eqref{DetofL}) is in turn equivalent to the so-called {\em optical constraint}~\cite{OrtPraPra09,OrtPraPra13}
\begin{align}\label{OpticalConstraint}
    L_{ik}L_{jk} = \alpha L_{(ij)} ,
\end{align}
as can be seen by rewriting~\eqref{OpticalConstraint} as $\frac{\alpha}{2}\left[(L^{-1})^T+L^{-1}\right]=\mathbbm{1}$ and using~\eqref{OpticalMatrixInverse} ($\alpha$ is a spacetime function which can be fixed by tracing the previous equation; the superscript $T$ denotes the transpose of a matrix).

In addition, eq.~\eqref{Sym_b} with \eqref{OpticalMatrixInverse} enables us to shift $r$ to set $b_{(ij)}=0$, such that from now on we have
\begin{align}
\label{NewOpticalMatrixInverse}
    b^{T}=-b . 
\end{align} 
Note that the antisymmetric matrix $b$ thus describes the leading order term (for large $r$) of the twist matrix of $\bl$, which is defined \cite{Pravdaetal04,OrtPraPra07} as $L_{[ij]}$. Throughout the paper, we will restrict the analysis to the case $b\neq0$ -- the case $b=0$ reduces to the (shearfree, twistfree) Robinson-Trautman metrics, fully analyzed in \cite{PodOrt06}.

For powers of the matrix $b$ we shall use the compact notation
\begin{align}
    b^s_{ij} \equiv \underbrace{b_{ik}b_{kl}\dots b_{mj}}_\text{$s$-times} \equiv \left( b^s \right)_{ij} ,
		\label{bij_power}
\end{align}
not to be confused with powers of specific matrix components, denoted as
\begin{align}
    b^s_{23} \equiv \left( b_{23} \right)^s.
		\label{b23_power}
\end{align}
For later calculations, it is useful to note that, thanks to~\eqref{NewOpticalMatrixInverse}, one has $b^s_{ij}=b^s_{ji}$ if $s$ is even and $b^s_{ij}=-b^s_{ji}$ if $s$ is odd.

\subsection{Ricci rotation coefficients and derivative operators}

\label{subsec_ricci_coeff}

We have specified some Ricci rotation coefficients in the preceding section, namely given by the equations \eqref{GeodandAffineCondition}, \eqref{ParaFrame} and \eqref{Li1} and also the result for the optical matrix, i.e., \eqref{OpticalMatrixInverse}, \eqref{NewOpticalMatrixInverse}. Now, we can fix the $r$-dependence of all the Ricci rotation coefficients and the frame vectors\footnote{More precisely, their $r$-dependence is fixed (except for \eqref{Ni1}) after we find the $r$-dependence of $\Phi_{ij}$ in the appendix~\ref{NPeqsPhi} in six dimensions, which is the main interest of our present investigation.}. Namely, (A.6, \cite{Durkeeetal10}) gives
\begin{align} \label{N_ij}
    N_{ij} = L_{mj}\left[ n_{im}-\int{L^{-1}_{nm}\Phi_{ni} \myd r}-\lambda \left( \frac{r^2}{2}\delta_{im}-rb_{im} \right) \right].
\end{align}
The equation (A.7, \cite{Durkeeetal10}) gives 
\begin{align}\label{Ni1}
    N_{i1} = \int{\mypsi_i \myd r} + n_{i1},
\end{align}
and identities (11b, 11n, 11a, and 11m)\footnote{Where for (11a) we used also (B.5, \cite{Pravdaetal04}).} of \cite{OrtPraPra07} give
\begin{align} \label{RicciRotCoeff_integrated}
    L_{1i} = L_{ji}l_{1j}, \quad \overset{i}{M}_{jk}=\overset{i}{m}_{jn}L_{nk}, \quad L_{11} = \frac{1}{n-2}L^{-1}_{ji}\Phi_{ij}+\lambda r + l_{11}, \quad \overset{i}{M}_{j1}=-2\int{\Phi^A_{ij} \myd r} + \overset{i}{m}_{j1}.
\end{align}

Let us write the basis vectors as
\begin{align}\label{BasisVectors}
    \bm{\ell} = \partial_r, \quad \mathbf{n} = U\partial_r + X^A\partial_A, \quad \mathbf{m}_{(i)} = \omega_i\partial_r + \xi^A_i\partial_A ,
\end{align}
where $U$, $X^A$, $\omega_i$ and $\xi^A_i$ are spacetime functions (to be determined in the following), $\pa_A=\pa/\pa x^A$, and the $x^A$ represent any set of ($n-1$) scalar functions such that $(r,x^A)$ is a well-behaved coordinate system. From the commutators \eqref{B.12}--\eqref{B.15} it follows
\begin{align}
		\label{MetricFunctions}
    &\omega_i = -L_{1i}r+L_{ji}\omega^0_j, \quad \xi^A_i = \xi^{A0}_jL_{ji} \nonumber \\
    &X^A=X^{A0}, \quad U = -l_{11}r-\frac{1}{n-2}\int{L^{-1}_{ji}\Phi_{ij} \myd r}+U^0-\dfrac{\lambda}{2} r^2.
\end{align}

For later purposes, let us note that the inverse metric can be written in terms of the above quantities as
\begin{align}\label{InverseMetric_General}
    \bm{g}^{-1} = (2U+\omega_i \omega_i)\partial_r^2+2(X^{A0}+\omega_i \xi_i^A) \partial_r \partial_A+ \xi_i^A \xi_i^B \partial_A \partial_B .
\end{align}

\subsection{Weyl scalars of zero boost weight}

\label{subsec_bw0}

Now, we will use the approach using expansions\footnote{Note that in the particular case $n=6$ relevant to the present paper, the forthcoming asymptotic behaviour follows also from the corresponding expressions obtained in closed form in appendix~\ref{NPeqsPhi}. The $r$-dependence of all quantities \eqref{N_ij}, \eqref{RicciRotCoeff_integrated}--\eqref{MetricFunctions}, and therefore also of the metric, is thus also known in closed form in six dimensions, see eq.~\eqref{InverseMetricin6D_General} below.\label{footn_6D}} similarly as in \cite{Ortaggio17} in order to obtain various constraints between asymptotic Ricci rotation coefficients and the functions \eqref{MetricFunctions} by examining the Newman--Penrose equations (the frame-projected Bianchi and Ricci equations). From now on, we take into account the following expansion of $L_{ij}$ \cite{OrtPraPra09b,Ortaggio17}, which follows from \eqref{OpticalMatrixInverse}
\begin{align}
    L_{ij} = \dfrac{\delta_{ij}}{r}+\dfrac{b_{ij}}{r^2}+\dfrac{b^2_{ij}}{r^3}+\dfrac{b^3_{ij}}{r^4}+O(r^{-5}) ,
\end{align}
where now $b_{ij}=b_{[ij]}$ (cf.~\eqref{NewOpticalMatrixInverse}).

When examining the NP equations of appendix~\ref{app_NP}, it is necessary to use also higher orders of the boost weight (b.w.)~0 Weyl scalars $\Phi_{ij}$ and $C_{ijkm}$, given by (as follows from \eqref{bw0DiffBianchiPhi}, \eqref{bw0DiffBianchiC}, cf. \cite{OrtPraPra09b,OrtPra14,Ortaggio17})
\begin{align}
\label{b.w.0WeylScalars_Asymptotics}
    \Phi_{ij} &= \Phi_0\left[ (n-3) \dfrac{\delta_{ij}}{r^{n-1}}+(n-1)\dfrac{b_{ij}}{r^n}+\dfrac{n-1}{2}\dfrac{\mathrm{Tr}b^2\delta_{ij}+2b^2_{ij}}{r^{n+1}} + \dfrac{n+1}{2}\dfrac{\mathrm{Tr}b^2b_{ij}+2b^3_{ij}}{r^{n+2}} \right]+O(r^{-(n+3)}), \nonumber \\
    C_{ijkm}&=4\Phi_0\left[ \dfrac{\delta_{i[m}\delta_{k]j}}{r^{n-1}}-\dfrac{2b^2_{[m|[j}\delta_{i]|k]}+\frac{1}{2}\mathrm{Tr}b^2\delta_{i[k}\delta_{m]j}+b_{i[k}b_{m]j}-b_{ij}b_{km}}{r^{n+1}} \right] + O(r^{-(n+3)}) ,
\end{align}
where $\Phi_0\neq0$ is an integration function independent of $r$ (the excluded case $\Phi_0=0$ simply corresponds to a spacetime of constant curvature, cf.~\cite{Pravdaetal04,OrtPraPra09b,OrtPra14,deFGodRea15,Ortaggio17} and proposition~1 of \cite{OrtPraPra18}; note that $\Phi_0$ differs by a numerical factor from the corresponding quantity used in \cite{Ortaggio17}). 
Terms of higher order are not needed for our purposes, but they can be determined recursively to any desired order once the leading terms~\eqref{b.w.0WeylScalars_Asymptotics} are known, and do {\em not} involve any integration functions other than $\Phi_0$ and $b_{ij}$. Hence, the full spacetime metric is {\em uniquely} determined by knowing $b_{ij}$, $l_{1i}$, $l_{11}$, $\omega^0_i$, $\xi^{A0}_i$, $X^{A0}$, $U^0$ and $\Phi_0$, as follows from \eqref{InverseMetric_General} with \eqref{OpticalMatrixInverse} and \eqref{MetricFunctions}.

With no loss of generality, from now on we shall take
\be
 \Phi_0 = \text{const}\neq 0 ,
 \label{Phi_const}
\ee
which amounts to an $r$-independent rescaling of $r$ and a corresponding rescaling (boost) of $\bl$ \cite{Kinnersley69,deFGodRea15,Ortaggio17}, cf. also a comment at the end of appendix~\ref{app_Phi_6D}.

\subsection{Weyl scalars of negative boost weight}

The Bianchi identities \eqref{B.1} and \eqref{B.6} give
\begin{align}
    \mypsi_{ijk} = \dfrac{2}{n-2}\mypsi_{[j}\delta_{k]i} + O(r^{-n}), \qquad \mypsi_i = -(n-1)(n-2)\dfrac{\Phi_0 l_{1i}}{r^{n-1}}+O(r^{-n}),
\end{align}
while from $\mypsi_i = \mypsi_{kik}$ it follows
\begin{align}\label{l1i_vanishing}
    l_{1i} = 0 , 
\end{align}
and thus $L_{1i} = 0$ and $\mypsi_{ijk} = O(r^{-n})$.

Next, from the Ricci identities~\eqref{11k} and \eqref{11i} we get
\begin{align}
    \xi^{A0}_{[j|}b_{i|k],A} &= \omega^0_{[j}\delta_{k]i} + b_{il}\overset{l}{m}_{[jk]}+b_{l[j|}\overset{l}{m}_{i|k]}, \label{xi on bij antisymmetric from Ricci}\\
    \label{eqfor_nij}
    X^{A0}b_{ij,A} &= U^0\delta_{ij}-n_{ij}-l_{11}b_{ij}-b_{kj}\overset{k}{m}_{i1}-b_{ik}\overset{k}{m}_{j1}-\lambda b^2_{ij}.
\end{align}
Using \eqref{xi on bij antisymmetric from Ricci}, from \eqref{B.1} and \eqref{B.9} it follows
\begin{align} \label{negative bw Weyl - final asymptotic behaviour}
    \mypsi_{ijk} = \dfrac{2}{n-3}\mypsi_{[j}\delta_{k]i} + O(r^{-(n+1)}), \quad \mypsi_i = (n-1)(n-3)\frac{\Phi_0 \omega^0_i}{r^n}+ O(r^{-(n+1)}), 
\end{align}
where the first of \eqref{negative bw Weyl - final asymptotic behaviour} compared with \eqref{B.6} gives
\begin{align}\label{Xi_bijA}
    \xi^{A0}_ib_{jk,A}=2\omega^0_{[j}\delta_{k]i}+2b_{l[j}\overset{l}m_{k]i}.
\end{align}
Next, \eqref{B.4} is identically satisfied at the order $O(r^{-(n-3)})$ and it reduces to
\begin{align}
    D(\myomega_{ik}L^{-1}_{kj}) =(n-1)(n-3)\Phi_0l_{11}\delta_{ij}r^{-n+2}+O(r^{-n+1}) . 
\end{align}

However, \eqref{B.13} reveals that 
\begin{align}\label{l11Vanish}
    l_{11} = 0 ,
\end{align}
and therefore the leading order of $\myomega_{ij}$ is $O(r^{-n+1})$. Hence, from the order $O(r^{-n+1})$ of \eqref{B.4} and $\myomega_{[ij]}=0$ we get
\begin{align}
    n_{[ij]}=0,
\end{align}
and so from (the symmetric and antisymmetric parts of) \eqref{eqfor_nij} it follows
\begin{align}\label{nij}
    &n_{ij} = n_{(ij)} = U^0\delta_{ij}-\lambda b^2_{ij}, \\ \label{bij,AConstraint}&X^{A0}b_{ij,A}=2b_{k[i}\overset{k}{m}_{j]1}. 
\end{align}
This in turn gives $\myomega_{ij} = O(r^{-n})$. At the next order, comparing \eqref{B.4} with the trace of \eqref{B.13}, one gets
\begin{align}\label{XiOmegaASCondition}
    &\xi^{A0}_{[i}\omega^0_{j],A}=2U^0b_{ij}+\omega^0_k\overset{k}{m}_{[ij]}-\lambda b^3_{ij},\\ \label{XiOmegaSymCondition}
    &\xi^{A0}_{(i}\omega^0_{j),A}+\omega^0_k\overset{k}{m}_{(ij)}=0.
\end{align}
Now, the anti-symmetric part of $\myomega_{ij}$ of the order $O(r^{-n})$ reduces to \eqref{XiOmegaASCondition} (thanks to $\myomega_{[ij]}=0$), while its symmetric part also vanishes thanks to \eqref{XiOmegaSymCondition}. Therefore, we have 
\begin{align}
    \myomega_{ij} = O(r^{-(n+1)}).
\end{align}

As in the case of b.w.~0 Weyl components in section~\ref{subsec_bw0}, also the components $\mypsi_{ijk}$, $\mypsi_i$ and $\myomega_{ij}$ can be determined recursively to any desired order, without involving any new integration functions. Again, their explicit form will not be needed in what follows.

In order to obtain further conditions from the commutators and the Ricci identities in the next section, it is now useful to summarize the $r$-dependence of the Ricci rotation coefficients and of the derivative operators obtained above (to the orders needed in the following), namely
\beqn
  & & L_{1i}=0 , \qquad \M{i}{j}{k}=\frac{\m{i}{j}{k}}{r}+\frac{\m{i}{j}{n}b_{nk}}{r^2}+\OO{3} , \qquad L_{11}=\lambda r+(n-3)\frac{\Phi_0}{r^{n-2}}+\OO{n} , \label{ricci_asym1} \\
	& & N_{ij}=\frac{\lambda}{2}(-r\delta_{ij}+b_{ij})+\frac{U^0\delta_{ij}-\frac{\lambda}{2}b^2_{ij}}{r}+\frac{U^0b_{ij}-\frac{\lambda}{2}b^3_{ij}}{r^2}+\OO{3} , \\
	& & \M{i}{j}{1}=\m{i}{j}{1}+\OOp{1-n} ,  \qquad N_{i1}=n_{i 1}+\OOp{1-n} , 
\eeqn
\beqn
 & & \omega_i=\frac{\omega^0_i}{r}+\frac{b_{ji}\omega^0_j}{r^2}+\OO{3} , \qquad  \xi^A_i=\frac{\xi^{A0}_i}{r}+\frac{b_{ji}\xi^{A0}_j}{r^2}+\OO{3}  , \\
 & & X^A=X^{A0} , \qquad U=-\frac{\lambda}{2}r^2+U^0+\frac{\Phi_0}{r^{n-3}}+\OOp{1-n} . 
\eeqn

\subsection{Further conditions from commutators and Ricci identities}

The commutator \eqref{B.14} applied on $r$ gives at the leading and subleading orders, respectively, 
\beqn
    & & n_{i1} = -\lambda \omega^0_{i} , \label{ni1} \\
    & & \xi^{A0}_iU^0_{,A}-X^{A0}\omega^0_{i,A}=(\overset{j}{m}_{i1}+2\lambda b_{ji})\omega^0_j . \label{XiUAminusXomega}
\eeqn

Next, \eqref{B.14} applied on $x^A$ gives at the leading order 
\begin{align}\label{xiB0XA0_minus_XB0xiAO}
     &\xi_i^{B0} X^{A0}_{,B}-X^{B0}\xi_{i ,B}^{A0} = (\overset{j}{m}_{i1}+\lambda b_{ji})\xi_j^{A0}.
\end{align}

At the leading order, \eqref{B.15} applied on $x^A$ gives
\begin{align}\label{xiB0xiA0}
    \xi_{[i}^{B0}\xi^{A0}_{j],B} = X^{A0}b_{ij}+\xi^{A0}_k\overset{k}{m}_{[ij]},
\end{align}
while applying it on $r$ at the leading order gives again \eqref{XiOmegaASCondition}.

The leading and subleading orders (i.e., $O(r^2)$ and $O(r)$) of the Ricci identity~\eqref{11h} turn out to be satisfied identically, while $O(1)$ reduces to \eqref{bij,AConstraint}. The symmetric part of the first non-trivial order gives (using \eqref{XiOmegaSymCondition} and \eqref{bij,AConstraint}) 
\begin{align}\label{XUAVanish}
    X^{A0}U^0_{,A}=0 ,
\end{align}
and its anti-symmetric part is identically satisfied thanks to \eqref{XiOmegaASCondition}.

Next, the first non-trivial order of \eqref{11l} gives (using \eqref{Xi_bijA}, \eqref{nij}, \eqref{ni1}) 
\begin{align}\label{XiA0UA_equals_btimesomega}
    \xi^{A0}_iU^0_{,A}=\lambda b_{ji}\omega^0_j.
\end{align}

The latter enables one to rewrite \eqref{XiUAminusXomega} as
\begin{align}\label{XAOmega0iA_plus_mjiOmega0j}
    X^{A0}\omega^0_{i,A} =-(\overset{j}{m}_{i1}+\lambda b_{ji})\omega^0_j .
\end{align}

Subsequently, the leading order of \eqref{11o} is identically satisfied, while the subleading order gives
\begin{align}\label{XAmijk,A_minus_xiA0mij1,A}
    X^{A0}\overset{i}{m}_{jk,A}-\xi^{A0}_k\overset{i}{m}_{j1,A}=-2\overset{l}{m}_{[i|1}\overset{l}{m}_{|j]k}-(\overset{l}{m}_{k1}+\lambda b_{lk})\overset{i}{m}_{jl}+2\lambda \omega^0_{[i}\delta_{j]k}.
\end{align}

Finally, the first non-trivial order of \eqref{11p} gives
\begin{align}\label{xiA0kmijl,A}
    \xi^{A0}_{[k|}\overset{i}{m}_{j|l],A} = 2U^0\delta_{i[l}\delta_{k]j}+b_{kl}\overset{i}{m}_{j1}-\overset{i}{m}_{n[k|}\overset{j}{m}_{n|l]}+\overset{i}{m}_{jn}\overset{n}{m}_{[kl]}-\lambda b_{i[l}b_{k]j}+\lambda(b^2_{i[k}\delta_{l]j}-b^2_{j[k}\delta_{l]i}). 
\end{align}

\section{Adapted coordinates and choice of a preferred frame}

\label{sec_adapted}

\subsection{Coordinates $(r,u)$ in the 2-plane spanned by $\bl$ and $\bn$}

\label{subsec_coord_u}

Let us recall that $\bl=\pa_r$ and $\bn=U\pa_r+X^A\pa_A$ (eq.~\eqref{BasisVectors}). Furthermore, we had that $X^A=X^{A0}$ is independent of $r$ (eq.~\eqref{MetricFunctions}), which means 
\begin{align}
    [  \partial_r, X^A\partial_A ] = 0 .
\end{align}
These two vector fields can thus be a part of a holonomic basis, i.e., we can define a coordinate $u$ such that $X^A\partial_A = \partial_u$. Therefore, we have now a coordinate system $(r, x^A) \equiv (r, u, x^\alpha)$ ($\alpha$ labels the remaining ($n-2$) coordinates, which will be specified subsequently). Hence, 
\begin{align}\label{XA_equals_deltaAu}
    X^A=\delta^A_u,
\end{align}
such that
\begin{align}
    \bm{n} = U\partial_r+\partial_u.
\end{align}

Thanks to this choice, equations~\eqref{bij,AConstraint}, \eqref{xiB0XA0_minus_XB0xiAO}, \eqref{XAOmega0iA_plus_mjiOmega0j}, \eqref{XAmijk,A_minus_xiA0mij1,A} and \eqref{xiB0xiA0} reduce to, respectively, 
\begin{align}
    &b_{ij,u}=2b_{k[i}\overset{k}{m}_{j]1}, \label{bij,u} \\
    &\xi_{i ,u}^{A0} = -(\overset{j}{m}_{i1}+\lambda b_{ji})\xi_j^{A0}, \label{xiA0_i,u} \\
    &\omega^0_{i,u} = -(\overset{j}{m}_{i1}+\lambda b_{ji})\omega^0_j, \label{omega0_i,u} \\
		&\overset{i}{m}_{jk,u}-\xi^{A0}_k\overset{i}{m}_{j1,A}=-2\overset{l}{m}_{[i|1}\overset{l}{m}_{|j]k}-(\overset{l}{m}_{k1}+\lambda b_{lk})\overset{i}{m}_{jl}+2\lambda \omega^0_{[i}\delta_{j]k}, \label{mi_jk,u}\\
    &\xi_{[i}^{B0}\xi^{u0}_{j],B} = b_{ij}+\xi^{u0}_k\overset{k}{m}_{[ij]}, \qquad  \xi_{[i}^{B0}\xi^{\alpha 0}_{j],B} =\xi^{\alpha 0}_k\overset{k}{m}_{[ij]} , \label{xiB0_xiu0}
\end{align}
while (by~\eqref{XUAVanish} and \eqref{XA_equals_deltaAu}) $U^0$ is independent of $u$, so that~\eqref{XiA0UA_equals_btimesomega} reads
\begin{align} \label{XiAlpha_on_U0}
        \xi^{\alpha0}_iU^0_{,\alpha}=\lambda b_{ji}\omega^0_j.
\end{align}

\subsection{Choice of the vectors $\bm{m}_{(i)}$ (six dimensions)}

\label{subsec_mi}

{\em From now on, we focus on the $n=6$ case.} A remaining freedom of $r$-independent spatial rotations can be used \cite{OrtPraPra07,Ortaggio17} (cf. also \cite{OrtPraPra09,OrtPraPra10}) to adapt the spacelike frame vectors to an eigenframe of the antisymmetric matrix $b_{ij}$, such that hereafter
\begin{align}\label{Lin6D}
    L = \mathrm{diag}\left( \dfrac{1}{r^2+b_{23}^2}\begin{bmatrix} r & b_{23} \\ -b_{23} & r \end{bmatrix}, \dfrac{1}{r^2+b_{45}^2}\begin{bmatrix} r & b_{45} \\ -b_{45} & r  \end{bmatrix} \right) .
\end{align}

Using this, the l.h.s and the r.h.s of~\eqref{bij,u} vanish separately, i.e., 
\begin{align}\label{bij,uVanish}
    b_{ij,u} = 0, \qquad b_{k[i}\overset{k}{m}_{j]1}=0	\quad (\mbox{in the frame~\eqref{Lin6D}}). 
\end{align}

Let us note that for $n=6$ we have obtained the explicit $r$-dependence of all Ricci rotation coefficients except for \eqref{Ni1}, and also of all of the derivative operators \eqref{BasisVectors} (see also footnote~\ref{footn_6D} and appendix~\ref{NPeqsPhi}). From this with~\eqref{InverseMetric_General}, \eqref{MetricFunctions}, \eqref{Lin6D} and \eqref{U}, the exact $r$-dependence of the metric reads 
\begin{align}\label{InverseMetricin6D_General}
     &\bm{g}^{-1} = \left(2U^0+\dfrac{2\Phi_0 r}{(r^2+b_{23}^2)(r^2+b_{45}^2)}-\lambda r^2+\dfrac{(\omega^0_2)^2+(\omega^0_3)^2}{r^2+b_{23}^2}+\dfrac{(\omega^0_4)^2+(\omega^0_5)^2}{r^2+b_{45}^2}\right)\partial_r^2+ 2\partial_r \partial_u \nonumber \\ & {}+ 2 \left(\dfrac{\omega^0_2 \xi^{A0}_2+\omega^0_3 \xi^{A0}_3}{r^2+b_{23}^2}+\dfrac{\omega^0_4 \xi^{A0}_4+\omega^0_5 \xi^{A0}_5}{r^2+b_{45}^2} \right)\partial_r \partial_A + \left(  \dfrac{\xi^{A0}_2 \xi^{B0}_2+\xi^{A0}_3 \xi^{B0}_3}{r^2+b_{23}^2}+\dfrac{\xi^{A0}_4 \xi^{B0}_4+\xi^{A0}_5 \xi^{B0}_5}{r^2+b_{45}^2} \right)\partial_A \partial_B \qquad (n=6) ,
\end{align}
where $A,B \in \{u, \alpha \}$.

Let us further observe that there is a scaling freedom (cf. also \cite{Ortaggio17}) which leaves the metric invariant and will be useful in the following. Namely, the coordinate transformation
\begin{align} \label{rescaling-begin}
    r^\prime = \kappa^{-1}r, \qquad u^\prime = \kappa u, 
\end{align}
where $\kappa \neq 0$ is a constant, accompanied by a boost
\begin{align}
    \bm{\ell}^\prime = \kappa \bm{\ell} = \partial_{r^\prime}, \qquad \bm{n}^\prime = \kappa^{-1}\bm{n} = \kappa^{-2}U\partial_{r^{\prime}}+\partial_{u^\prime},  
\end{align}
produces the rescaling 
\begin{align} \label{rescaling-end}
    b_{ij}^\prime = \kappa^{-1}b_{ij}, \quad U^{0\prime} = \kappa^{-2}U^0, \quad \Phi_0^\prime = \kappa^{-5}\Phi_0, \quad \omega_i^{0\prime} = \kappa^{-2}\omega_i^0, \quad \xi_i^{u0\prime} = \xi_i^{u0}, \quad \xi_i^{\alpha 0 \prime} = \kappa^{-1} \xi_i^{\alpha 0} . 
\end{align}

Similarly as in \cite{Ortaggio17}, the second of \eqref{bij,uVanish} needs to be investigated in two different cases, as we now discuss.

\begin{enumerate}[(i)]

\item\label{b_generic} Generic case $b_{45}\neq \pm b_{23}$: 
The second of~\eqref{bij,uVanish} means that the antisymmetric matrices $b_{ij}$ and $\overset{i}{m}_{j1}$ commute, therefore (since $b_{ij}$ is already block-diagonal; cf., e.g., cap.~IX of \cite{Gantmacherbook})
\begin{align}
    \overset{2}{m}_{41}=\overset{2}{m}_{51}=\overset{3}{m}_{41}=\overset{3}{m}_{51}=0,
\end{align}
so the only non-vanishing components of $\overset{i}{m}_{j1}$ are $\overset{2}{m}_{31}$, $\overset{4}{m}_{51}$. However, $\m{i}{j}{1}$ does {\em not} transform homogeneously under an $r$-independent but $u$-dependent rotation. Namely, a rotation in the plane $(23)$ by an angle $\theta$ leaves $b_{ij}$ unchanged, while (cf. \cite{OrtPraPra07} and eqs.~(3.12,3.13) of \cite{Ortaggio17})
\begin{align}
    \overset{2}{m}_{31}\mapsto\overset{2}{m}_{31}+\theta_{,u} .
\end{align}
This can be used to set $\overset{2}{m}_{31} = -\lambda b_{23}$. One can proceed similarly with $\overset{4}{m}_{51}$ and thus arrive at a frame such that $\overset{i}{m}_{j1}=-\lambda b_{ij}$.

\item\label{b_deg} Special case $b_{45}=b_{23}\neq 0$: in this case the eigenvalues of $b$ are degenerate and from~\eqref{bij,uVanish} one obtains only the weaker condition
\begin{align}
    \overset{2}{m}_{51} = -\overset{3}{m}_{41}, \qquad \overset{2}{m}_{41} = \overset{3}{m}_{51}.
\end{align}
However, the canonical form~\eqref{Lin6D} is now invariant under a larger set of spins. Proceeding as in \cite{Ortaggio17} (to which we refer for more details) this enables one to again arrive at $\overset{i}{m}_{j1}=-\lambda b_{ij}$. Let us note that in this case $\bl$ is twisting but {\em shearfree} (i.e., the symmetric part of $L$, eq.~\eqref{Lin6D}, is proportional to the identity matrix). This encompasses also the case $b_{45}=-b_{23}$, up to relabeling the frame vectors.

\end{enumerate}

In both the above cases, we can thus choose a parallelly transported frame such that
\begin{align}\label{mij1_prop_bij}
    \overset{i}{m}_{j1}+\lambda b_{ij}=0 .
\end{align}
Using the latter, eqs.~\eqref{omega0_i,u} and \eqref{xiA0_i,u} reduce to
\begin{align} 
\label{OmegaandXi_vanishinguderivative}
   \omega^0_{i,u} =0, \qquad \xi^{A0}_{i,u} = 0,
\end{align}
while \eqref{mi_jk,u} (with \eqref{Xi_bijA}) and \eqref{xiA0kmijl,A} give 
\begin{align}
    &\overset{i}{m}_{jk,u}=0, \\
    &\xi^{A0}_{[k|}\overset{i}{m}_{j|l],A} = 2U^0\delta_{i[l}\delta_{k]j}-\overset{i}{m}_{n[k|}\overset{j}{m}_{n|l]}+\overset{i}{m}_{jn}\overset{n}{m}_{[kl]}-\lambda b_{kl}b_{ij}-\lambda b_{i[l}b_{k]j}+\lambda(b^2_{i[k}\delta_{l]j}-b^2_{j[k}\delta_{l]i}).  \label{xiA_k_mijkl,A_inAdaptedCoordinates}
\end{align}

Next, eq.~\eqref{Xi_bijA} does not explicitly contain $\lambda$, so its various components (with \eqref{Lin6D}) read as in \cite{Ortaggio17}, i.e., 
\begin{align}
    \omega^0_2 &= \xi_3^{A0}b_{23,A} = b_{23}\overset{3}{m_{44}} + b_{45}\overset{2}{m_{54}} = b_{23}\overset{3}{m_{55}} - b_{45}\overset{2}{m_{45}}, \label{Appendix_omega2_generally}\\
    \omega^0_3 &= -\xi_2^{A0}b_{23,A} = -b_{23}\overset{2}{m_{44}} + b_{45}\overset{3}{m_{54}} = -b_{23}\overset{2}{m_{55}} - b_{45}\overset{3}{m_{45}},  \label{Appendix_omega3_generally}\\
    \omega^0_4 &= \xi_5^{A0}b_{45,A} = -b_{23}\overset{3}{m_{42}} - b_{45}\overset{2}{m_{52}} = b_{23}\overset{2}{m_{43}} - b_{45}\overset{3}{m_{53}},  \label{Appendix_omega4_generally}\\
    \omega^0_5 &= -\xi_4^{A0}b_{45,A} = -b_{23}\overset{3}{m_{52}} + b_{45}\overset{2}{m_{42}} = b_{23}\overset{2}{m_{53}} + b_{45}\overset{3}{m_{43}},  \label{Appendix_omega5_generally}\\
    \xi_4^{A0}b_{23,A} &= 0 = \xi_5^{A0}b_{23,A}, \qquad   \xi_2^{A0}b_{45,A} = 0 = \xi_3^{A0}b_{45,A}, \label{Appendix_bij_VanishingXiDerivatives}\\
    b_{45}\overset{2}{m_{53}} &= -b_{23}\overset{3}{m_{43}}, \qquad b_{45}\overset{3}{m_{52}} = b_{23}\overset{2}{m_{42}}, \qquad b_{45}\overset{2}{m_{43}} = b_{23}\overset{3}{m_{53}}, \qquad b_{45}\overset{3}{m_{42}} = -b_{23}\overset{2}{m_{52}}, \label{Appendix_Constraints_on_mijk_first}\\
    b_{45}\overset{2}{m_{55}} &= -b_{23}\overset{3}{m_{45}}, \qquad b_{45}\overset{2}{m_{44}} = b_{23}\overset{3}{m_{54}}, \qquad b_{45}\overset{3}{m_{55}} = b_{23}\overset{2}{m_{45}}, \qquad b_{45}\overset{3}{m_{44}} = -b_{23}\overset{2}{m_{54}}. \label{Appendix_Constraints_on_mijk_second}
\end{align}

The various conditions obtained above are the starting point in order to specify the dependence of all the metric functions also on the remaining coordinates $x^\alpha$, thus arriving at a final, canonical form of the metric. Further consequences of \eqref{Appendix_omega2_generally}--\eqref{Appendix_Constraints_on_mijk_second} need to be studied separately in the two possible cases~\eqref{b_generic} 
and \eqref{b_deg} defined above. The former will however suffice for the purposes of the present paper, as we discuss in the next section.

\section{Complete integration}

\label{sec_generic}

The genericity assumption~\ref{ass2b}. of section~\ref{subsec_summary} implies that $b_{45}^2\neq b_{23}^2$, which will be thus assumed hereafter.

\subsection{Obtaining the metric functions}

\label{subsec_general_sol}

Since $\omega^0_{i,u}=0=\omega^0_{i,r}$ (see \eqref{OmegaandXi_vanishinguderivative}), a residual freedom of $r$- and $u$-independent spins in the planes $(23)$ and $(45)$ can be used to set
\begin{align} \label{Omega3andOmega5Vanishes}
    \omega^0_3 = 0, \quad \omega^0_5=0 .
\end{align}

This reduces eqs.~\eqref{Appendix_omega2_generally}--\eqref{Appendix_Constraints_on_mijk_second} to (without loss of generality we assume $b_{45}\neq0$)
\begin{align}
    \omega^0_2 &= \xi_3^{A0}b_{23,A} = \dfrac{b_{23}^2-b_{45}^2}{b_{45}}\overset{2}{m_{45}}, \qquad \xi_2^{A0}b_{23,A} = \xi_4^{A0}b_{23,A} = \xi_5^{A0}b_{23,A} = 0, \label{Appendix_Omega2_subcaseb45_neq_pmb23}\\
     \omega^0_4 &= \xi_5^{A0}b_{45,A} =  \dfrac{b_{23}^2-b_{45}^2}{b_{45}}\overset{2}{m_{52}}, \qquad \xi_2^{A0}b_{45,A} = \xi_3^{A0}b_{45,A} = \xi_4^{A0}b_{45,A} = 0, \label{Appendix_Omega4_subcaseb45_neq_pmb23}\\
     \overset{3}{m_{53}} &= \overset{2}{m_{52}}, \qquad \overset{2}{m_{54}} = -\overset{2}{m_{45}}, \qquad -\overset{3}{m_{42}} = \overset{2}{m_{43}} = \dfrac{b_{23}}{b_{45}} \overset{2}{m_{52}}, \qquad \overset{3}{m_{55}} = \overset{3}{m_{44}} = \dfrac{b_{23}}{b_{45}} \overset{2}{m_{45}}, \label{Appendix_Constraints_on_mijk_subcaseb45_neq_pmb23_first}\\
     \overset{3}{m_{52}} &= \overset{2}{m_{42}} = \overset{2}{m_{53}} = \overset{3}{m_{43}} = \overset{2}{m_{44}} = \overset{3}{m_{54}} = \overset{2}{m_{55}} = \overset{3}{m_{45}} = 0. \label{Appendix_Constraints_on_mijk_subcaseb45_neq_pmb23_second}
\end{align}

Since $b_{ij}$ does not depend on $r$ and $u$, thanks to~\eqref{Appendix_Omega2_subcaseb45_neq_pmb23} and \eqref{Appendix_Omega4_subcaseb45_neq_pmb23} we note that
\begin{align}\label{b_and_omega_Equivalence}
    b_{23}=\text{const.} \Leftrightarrow \omega^0_2=0, \qquad   b_{45}=\text{const.} \Leftrightarrow \omega^0_4=0.
\end{align}

Next, from \eqref{XiOmegaSymCondition} with \eqref{XiOmegaASCondition} it follows (recall that~\eqref{XiOmegaASCondition} refers to the notation~\eqref{bij_power}, whereas~\eqref{U0b} to~\eqref{b23_power}) 
\begin{align}
    &\xi^{A0}_3\omega^0_{2,A} = -\omega^0_2\overset{2}{m_{32}}, \qquad \xi^{A0}_5\omega^0_{2,A} = \omega^0_4\overset{2}{m_{45}}, \qquad \xi^{A0}_2\omega^0_{2,A} = 0 = \xi^{A0}_4\omega^0_{2,A}, \label{Xi3onOmega2}\\
      &\xi^{A0}_5\omega^0_{4,A} = -\omega^0_4\overset{4}{m_{54}}, \qquad \xi^{A0}_3\omega^0_{4,A} = -\omega^0_2\overset{2}{m_{43}}, \qquad \xi^{A0}_4\omega^0_{4,A} = 0 = \xi^{A0}_2\omega^0_{4,A}, \label{Xi5onOmega4}\\
      &\omega^0_2\overset{2}{m_{33}} = 0, \qquad \omega^0_2\overset{2}{m_{35}} = 0, \qquad \omega^0_4\overset{4}{m_{55}} = 0, \qquad \omega^0_4\overset{4}{m_{53}} = 0, \label{Omega2m233}\\
      &\omega^0_2\overset{2}{m_{52}}+\omega^0_4\overset{4}{m_{52}} = 0, \qquad \omega^0_2\overset{2}{m_{34}} - \omega^0_4\overset{3}{m_{44}} = 0, \label{Omega2m252}\\
      &2U^0b_{23} = \omega^0_2\overset{2}{m_{32}} + \omega^0_4\overset{2}{m_{43}} - \lambda b^3_{23}, \quad 2U^0b_{45} = \omega^0_4\overset{4}{m_{54}} - \omega^0_2\overset{2}{m_{45}} - \lambda b^3_{45}. \label{U0b}
\end{align}

Using \eqref{OmegaandXi_vanishinguderivative}, \eqref{Appendix_Constraints_on_mijk_subcaseb45_neq_pmb23_first} and \eqref{Appendix_Constraints_on_mijk_subcaseb45_neq_pmb23_second}, the second of \eqref{xiB0_xiu0} results in
\begin{align}
    2\xi_{[2}^{\beta 0}\xi_{3],\beta}^{\alpha 0} &= -\xi_2^{\alpha 0} \overset{2}{m_{32}} - \xi_3^{\alpha 0} \overset{2}{m_{33}} - {2}\xi_4^{\alpha 0} \overset{2}{m_{43}}, \label{Appendix_Commutator_xi2xi3} \\
    2\xi_{[2}^{\beta 0}\xi_{4],\beta}^{\alpha 0} &= {2}\xi_3^{\alpha 0} \overset{3}{m_{[24]}} + {2}\xi_5^{\alpha 0} \overset{5}{{m_{[24]}}}, \\
    2\xi_{[2}^{\beta 0}\xi_{5],\beta}^{\alpha 0} &= -\xi_2^{\alpha 0} \overset{2}{m_{52}} - \xi_3^{\alpha 0} \overset{2}{m_{35}} + 2\xi_4^{\alpha 0} \overset{4}{m_{[25]}}, \\
    2\xi_{[4}^{\beta 0}\xi_{3],\beta}^{\alpha 0} &= \xi_4^{\alpha 0} \overset{3}{m_{44}} - \xi_5^{\alpha 0} \overset{4}{m_{53}} + 2\xi_2^{\alpha 0} \overset{2}{m_{[43]}}, \\
    2\xi_{[3}^{\beta 0}\xi_{5],\beta}^{\alpha 0} &= \xi_2^{\alpha 0} \overset{2}{m_{35}} - \xi_3^{\alpha 0} \overset{3}{m_{53}} - \xi_4^{\alpha 0} \overset{4}{m_{53}} - \xi_5^{\alpha 0} \overset{3}{m_{55}}, \\
    2\xi_{[4}^{\beta 0}\xi_{5],\beta}^{\alpha 0} &= 2\xi_2^{\alpha 0} \overset{2}{m_{45}} - \xi_4^{\alpha 0} \overset{4}{m_{54}} - \xi_5^{\alpha 0} \overset{4}{m_{55}}. \label{Appendix_Commutator_xi4xi5}
\end{align}

Applying these to $b_{23,\alpha}$, $b_{45,\alpha}$ and using \eqref{Appendix_Omega2_subcaseb45_neq_pmb23}, \eqref{Appendix_Omega4_subcaseb45_neq_pmb23} and \eqref{Xi3onOmega2}--\eqref{Omega2m233} gives
\begin{align}
    \omega_2^0\overset{3}{m_{[24]}} = 0, \qquad  \omega_2^0\overset{2}{m_{[34]}} = 0, \label{Appendix_omega2m324}\\
     \omega_4^0\overset{5}{m_{[24]}} = 0, \qquad  \omega_4^0\overset{4}{m_{[25]}} = 0.\label{Appendix_omega4m524}
\end{align}

To proceed with the analysis one thus needs to distinguish among various subcases, depending on the possible vanishing of  $\omega^0_2$ and $\omega^0_4$.
However, since in this paper we assume~\ref{ass2b}. of section~\ref{subsec_summary}, neither $b_{23}$ nor $b_{45}$ can be constant. Eq.~\eqref{b_and_omega_Equivalence} thus means that $\omega^0_2 \neq 0 \neq \omega^0_4$. Therefore, \eqref{Omega2m233}, \eqref{Appendix_omega2m324} and \eqref{Appendix_omega4m524} imply

\begin{align}
    \overset{2}{m}_{33} &= \overset{2}{m}_{35} = \overset{4}{m}_{55} = \overset{4}{m}_{53} = 0, \\
    \overset{3}{m}_{[24]} &= \overset{2}{m}_{[34]} = \overset{5}{m}_{[24]} = \overset{4}{m}_{[25]} = 0.
\end{align}
Using these, \eqref{Appendix_Commutator_xi2xi3}--\eqref{Appendix_Commutator_xi4xi5} reduce to (recall also~\eqref{Appendix_Constraints_on_mijk_subcaseb45_neq_pmb23_first})
\begin{align}\label{OldXiCommutators1}
    &2\xi^{\beta 0}_{[2}\xi^{\alpha 0}_{3], \beta} = -\xi^{\alpha 0}_2 \overset{2}{m_{32}}-2\xi^{\alpha 0}_4\overset{2}{m_{34}}, \qquad  2\xi^{\beta 0}_{[2}\xi^{\alpha 0}_{4], \beta} = 0, \qquad  2\xi^{\beta 0}_{[2}\xi^{\alpha 0}_{5], \beta} = -\xi^{\alpha 0}_2 \overset{2}{m_{52}} , \\ \label{OldXiCommutators2}
      &2\xi^{\beta 0}_{[4}\xi^{\alpha 0}_{3], \beta} = \xi^{\alpha 0}_4 \overset{3}{m_{44}}, \qquad 2\xi^{\beta 0}_{[3}\xi^{\alpha 0}_{5], \beta} = -\xi^{\alpha 0}_3 \overset{2}{m_{52}}-\xi^{\alpha 0}_5\overset{3}{m_{44}}, \qquad  2\xi^{\beta 0}_{[4}\xi^{\alpha 0}_{5], \beta} = 2\xi^{\alpha 0}_2 \overset{2}{m_{45}}-\xi^{\alpha 0}_4\overset{4}{m_{54}}.
\end{align}

These conditions can be used to construct four commuting vector fields spanning the subspace of the $x^\alpha$, namely (these do not contain $\lambda$ and thus coincide with the expressions given in~\cite{Ortaggio17}, cf. \cite{Kokoska_PhD} for more details)
\begin{align}
      &\hat{\bm{\xi}}_2 \equiv \left[ \mathcal \omega^0_2\left( \alpha b_{45}^2 + \beta \right)\xi^{\alpha 0}_2 + \omega^0_4\left( \alpha b_{23}^2 + \beta \right)\xi^{\alpha 0}_4 \right]\partial_\alpha, \qquad  \hat{\bm{\xi}}_3 \equiv  \dfrac{1}{\omega^0_2}\xi^{\alpha 0}_3\partial_\alpha, \label{xi2andxi3_final}\\ 
     &\hat{\bm{\xi}}_4 \equiv \left[ \mathcal \omega^0_2\left( \gamma b_{45}^2 + \delta \right)\xi^{\alpha 0}_2 + \omega^0_4\left( \gamma b_{23}^2 + \delta \right)\xi^{\alpha 0}_4 \right]\partial_\alpha, \qquad \hat{\bm{\xi}}_5 \equiv  \dfrac{1}{\omega^0_4}\xi^{\alpha 0}_5\partial_\alpha, \label{xi4andxi5_final}
\end{align}
where $\alpha,\beta,\gamma,\delta$ are constants which can be chosen arbitrarily provided $\alpha\delta-\beta\gamma\neq0$ (two convenient choices will be specified subsequently in~\eqref{gauge_EFCP} and below~\eqref{epsilon_Definition}, respectively).
Note that one further has $\left[ \hat{\bm{\xi}}_i, \partial_u \right] = 0 = \left[ \hat{\bm{\xi}}_i, \partial_r \right]$ for any $i, j=2,3,4,5$ (recall \eqref{bij,uVanish}, \eqref{OmegaandXi_vanishinguderivative}).
 Therefore, one can complete a coordinate basis by defining $x^\alpha = (\phi_1, y_1, \phi_2, y_2)$ via 
\begin{align} \label{TransverseCoordinates_Definition}
    \partial_{\phi_1} \equiv \hat{\bm{\xi}}_2, \qquad  \partial_{y_1} \equiv \hat{\bm{\xi}}_3, \qquad  \partial_{\phi_2} \equiv \hat{\bm{\xi}}_4, \qquad  \partial_{y_2} \equiv \hat{\bm{\xi}}_5 .
\end{align}
Recalling \eqref{Appendix_Omega2_subcaseb45_neq_pmb23} and \eqref{Appendix_Omega4_subcaseb45_neq_pmb23}, it follows \cite{Ortaggio17} that one can always choose $(y_1,y_2)$ such that
\begin{align}
    y_1 \equiv b_{23}, \qquad y_2 \equiv b_{45}.
\end{align}

Using these coordinates, from \eqref{XiAlpha_on_U0} with \eqref{Omega3andOmega5Vanishes} one gets
\begin{align} \label{U0_Final}
    2U^0 = 2\hat{\mathcal{U}}^0 + \lambda\left( y_1^2 + y_2^2 \right), 
\end{align}
where $\hat{\mathcal{U}}^0$ is an integration constant.

Next, eqs.~\eqref{Xi3onOmega2} and \eqref{Xi5onOmega4} imply that $\omega^0_2$ and $\omega^0_4$ do not depend on $(\phi_1,\phi_2)$, and further give
\begin{align} \label{CoordinateDerivativesofOmegas}
    \omega^0_{2,y_2}=\overset{2}{m_{45}}, \qquad \omega^0_{4,y_1}=-\overset{2}{m_{43}}, \qquad \omega^0_{2,y_1}=-\overset{2}{m_{32}}, \qquad \omega^0_{4,y_2}=-\overset{4}{m_{54}} .
\end{align}
The first two of~\eqref{CoordinateDerivativesofOmegas} (using a subset of \eqref{Appendix_Omega2_subcaseb45_neq_pmb23}--\eqref{Appendix_Constraints_on_mijk_subcaseb45_neq_pmb23_first}) lead to
\begin{align}  \label{Omega2andOmega4Reduced}
    \left( \omega^0_2 \right)^2 = \dfrac{\mathcal{G}(y_1)}{y_1^2-y_2^2}, \qquad \left( \omega^0_4 \right)^2 = \dfrac{\mathcal{H}(y_2)}{y_1^2-y_2^2},
\end{align}
where $\mathcal{G}$ and $\mathcal{H}$ are integration functions.
Then, the last two of~\eqref{CoordinateDerivativesofOmegas} (using \eqref{U0b}, a subset of \eqref{Appendix_Omega2_subcaseb45_neq_pmb23}--\eqref{Appendix_Constraints_on_mijk_subcaseb45_neq_pmb23_first}, \eqref{U0_Final} and \eqref{Omega2andOmega4Reduced}) reveal that $\left( \omega^0_2 \right)^2+\left( \omega^0_4 \right)^2=-2\hat{\mathcal{U}}^0(y_1^2+y_2^2)-\lambda(y_1^4+y_2^4+y_1^2y_2^2)+c_0$. Together with~\eqref{Omega2andOmega4Reduced}, one eventually concludes that
\be
\label{Omegas_Final}
    \left( \omega^0_2 \right)^2 = \frac{{\cal P}(y_1)}{y_2^2-y_1^2}, \qquad  \left( \omega^0_4 \right)^2 = \frac{{\cal P}(y_2)}{y_1^2-y_2^2}, \qquad {\cal P}(s)\equiv\lambda s^6 + 2\hat{\mathcal{U}}^0s^4 - c_0s^2 - d_0 ,
\ee
where $c_0$ and $d_0$ are integration constants.

Further, we can also invert~\eqref{xi2andxi3_final}, \eqref{xi4andxi5_final} (with \eqref{TransverseCoordinates_Definition}, \eqref{Polynomial completely factorized}, \eqref{2U0Definition}) to get the components $\xi^{\alpha0}_i$ in the form
\begin{align}
    &\xi_2^{\alpha 0}\partial_\alpha = \dfrac{\omega^0_2(\beta \gamma -\alpha \delta)^{-1}}{{\cal P}(y_1)}\left[ -(\gamma y_1^2+\delta)\partial_{\phi_1} + (\alpha y_1^2 + \beta)\partial_{\phi_2} \right], \label{xi20} \\
     &\xi_4^{\alpha 0}\partial_\alpha = \dfrac{\omega^0_4(\beta \gamma -\alpha \delta)^{-1}}{{\cal P}(y_2)}\left[ -(\gamma y_2^2+\delta)\partial_{\phi_1} + (\alpha y_2^2 + \beta)\partial_{\phi_2}\right], \\
     &\xi_3^{\alpha 0}\partial_{\alpha} = \omega^0_2\partial_{y_1}, \qquad  \xi_5^{\alpha 0}\partial_{\alpha} = \omega^0_4\partial_{y_2}. \label{xi50}
\end{align}

Let us finally determine the remaining components $\xi^{u 0}_{i}$ using the first of \eqref{xiB0_xiu0}, which is algebraically the same as in \cite{Ortaggio17}. First, its $ij=35$ component gives (cf. \cite{Ortaggio17} for some of the intermediate steps)
\begin{align} 
\label{xi3u0 and xi5u0 integrability condition}
    \left( \dfrac{\xi_3^{u0}}{\omega^0_2} \right)_{,y_2} =  \left( \dfrac{\xi_5^{u0}}{\omega^0_4} \right)_{,y_1} .
\end{align}
As in \cite{Ortaggio17}, this can be used  to set $\xi_3^{u0} = 0 = \xi_5^{u0}$ \cite{Ortaggio17}. Namely, the transformation 
\begin{align} \label{u-transformation doubly-spinning Kerr-(A)dS}
    u \mapsto u + V(\phi_1, \phi_2, y_1, y_2),
\end{align}
results in
\begin{align} \label{u-transformation doubly-spinning Kerr-(A)dS -- xiu0 transformations}
    \xi^{u0}_2 \mapsto \xi^{u0}_2 + \xi^{\alpha 0}_2 V_{,\alpha}, \qquad \xi^{u0}_3 \mapsto \xi^{u0}_3 + \omega^0_2 V_{,y_1}, \qquad \xi^{u0}_4 \mapsto \xi^{u0}_4 + \xi^{\alpha 0}_4 V_{,\alpha}, \qquad \xi^{u0}_5 \mapsto \xi^{u0}_5 + \omega^0_5 V_{,y_2} . 
\end{align}
One can choose $V$ such as to simultaneously set $\xi^{u0}_3 + \omega^0_2 V_{,y_1} = 0 = \xi^{u0}_5 + \omega^0_5 V_{,y_2}$, since the integrability condition is \eqref{xi3u0 and xi5u0 integrability condition} and is thus identically satisfied. From now on, we thus have 
\begin{align}
    \xi_3^{u0} = 0 = \xi_5^{u0} ,
		\label{xiu3_5=0}
\end{align}
which simplifies the remaining components of the first of \eqref{xiB0_xiu0}. The components $ij=25,23,34,45$  give 
\begin{align}
    \xi^{u0}_{2,y_2} &= -\dfrac{y_2}{y_2^2-y_1^2}\xi^{u0}_2, \qquad -\left( \omega^0_2 \xi^{u0}_2 \right)_{,y_1} = 2y_1 + 2\xi^{u0}_4 \omega^0_{4,y_1}, \\
    \xi^{u0}_{4,y_1} &= -\dfrac{y_1}{y_1^2-y_2^2}\xi^{u0}_4, \qquad -\left( \omega^0_4 \xi^{u0}_4 \right)_{,y_2} = 2y_2 + 2\xi^{u0}_2 \omega^0_{2,y_2}.
\end{align}
Integrating these yields
\begin{align} \label{xi2uandxi4u_Integrated}
    \xi^{u0}_2 = \dfrac{y_1^4 - e_0y_1^2-f_0}{{\cal P}(y_1)}\omega^0_2, \qquad \xi^{u0}_4 = \dfrac{y_2^4 - e_0y_2^2-f_0}{{\cal P}(y_2)}\omega^0_4,
\end{align}
where $e_0 = e_0(\phi_1, \phi_2)$ and $f_0 = f_0(\phi_1, \phi_2)$ are integration functions. These are, however, constrained by the $ij = 24$ component of the first of \eqref{xiB0_xiu0}, which now reads 
\begin{align} \label{e_0 and f_0 integrability condition}
    \left( e_0\delta - f_0\gamma \right)_{,\phi_1} = \left( e_0\beta - f_0\alpha \right)_{,\phi_2} .
\end{align}

Under a transformation \eqref{u-transformation doubly-spinning Kerr-(A)dS}, \eqref{u-transformation doubly-spinning Kerr-(A)dS -- xiu0 transformations} with $V_{,y_1} = 0 = V_{,y_2}$ (thus not affecting~\eqref{xiu3_5=0}), the integration functions in \eqref{xi2uandxi4u_Integrated} transform as
\begin{align}
    e_0 \mapsto e_0 + \dfrac{\gamma}{\beta \gamma - \alpha \delta}V_{,\phi_1} - \dfrac{\alpha}{\beta \gamma - \alpha \delta}V_{,\phi_2}, \qquad f_0 \mapsto f_0 + \dfrac{\delta}{\beta \gamma - \alpha \delta}V_{,\phi_1} - \dfrac{\beta}{\beta \gamma - \alpha \delta}V_{,\phi_2} ,
		\label{rescal_e0_f0}
\end{align}
which can be used to set both $e_0$ and $f_0$ to arbitrary {\em constant} values (since~\eqref{e_0 and f_0 integrability condition} means the corresponding integrability condition is satisfied). This will be useful in sections~\ref{subsubsec_general_metric} and \ref{subsubsec_recovering}.

\subsubsection{General metric in a convenient gauge}

\label{subsubsec_general_metric}

As a summary, we have obtained the asymptotic quantities given by~\eqref{Omega3andOmega5Vanishes}, \eqref{Omegas_Final}, \eqref{xi20}--\eqref{xi50}, \eqref{xi2uandxi4u_Integrated}, \eqref{U0_Final} and \eqref{XA_equals_deltaAu}. This fully specifies the contravariant metric~\eqref{InverseMetricin6D_General}. Using the remaining ``gauge'' freedom, a convenient choice of the constants $\alpha$, $\beta$, $\gamma$, $\delta$ and $e_0$, $f_0$ (see comments below~\eqref{xi4andxi5_final} and \eqref{rescal_e0_f0}, respectively) turns out to be
\be
  \alpha=0=\delta, \quad \beta=1=\gamma, \qquad e_0=0=f_0. 
	\label{gauge_EFCP}
\ee
After relabeling the constant $\Phi_0$ of \eqref{b.w.0WeylScalars_Asymptotics}, \eqref{Phi_const} as
\be
 \mu \equiv -2\Phi_0 ,
\ee
this results in the metric
\beqn
\label{CarterPlebanskiEF}
    & & \mathrm{d}s^2 = 2\mathrm{d}r\left[ \mathrm{d}u + \left( y_1^2+y_2^2 \right)\mathrm{d}\phi_1 + y_1^2y_2^2\mathrm{d}\phi_2 \right] + \left( r^2 + y_1^2 \right)\dfrac{y_2^2-y_1^2}{\mathcal{P}(y_1)}\mathrm{d}y_1^2 + \left( r^2 + y_2^2 \right)\dfrac{y_1^2-y_2^2}{\mathcal{P}(y_2)}\mathrm{d}y_2^2 \nonumber \\ 
		& & \qquad {}+ \dfrac{\mathcal{P}(y_1)}{(r^2+y_1^2)(y_2^2-y_1^2)}\left[ \mathrm{d}u + \left( y_2^2-r^2 \right)\mathrm{d}\phi_1 - r^2y_2^2\mathrm{d}\phi_2 \right]^2+ 
		\dfrac{\mathcal{P}(y_2)}{(r^2+y_2^2)(y_1^2-y_2^2)}\left[ \mathrm{d}u + \left( y_1^2-r^2 \right)\mathrm{d}\phi_1 -r^2y_1^2\mathrm{d}\phi_2 \right]^2 \nonumber \\ 
		& & \qquad {}+\dfrac{\mathcal{Q}(r)}{(r^2+y_1^2)(r^2+y_2^2)}\left[\mathrm{d}u + \left( y_1^2+y_2^2 \right)\mathrm{d}\phi_1 + y_1^2y_2^2\mathrm{d}\phi_2 \right]^2 ,
\eeqn
where the metric functions are defined by~\eqref{Omegas_Final} and 
\be
  {\cal Q}(r)\equiv\lambda r^6 -2\hat{\mathcal{U}}^0r^4 - c_0r^2 + \mu r + d_0  \label{Q_main} .
\ee

The Einstein spacetimes~\eqref{CarterPlebanskiEF} are of constant curvature iff $\mu=0$, which also implies that the metric is manifestly KS, where the KS covector is given by the mWAND $\ell_a\d x^a=\mathrm{d}u + \left( y_1^2+y_2^2 \right)\mathrm{d}\phi_1 + y_1^2y_2^2\mathrm{d}\phi_2$. The coordinate transformation~\eqref{transf_CP_EF} given in appendix~\ref{subsubsec_generalized} reveals that line-element~\eqref{CarterPlebanskiEF} (i.e., \eqref{CarterPlebanskiEF_app}) is locally isometric to a subfamily of the general Kerr-NUT-(A)dS family of \cite{CheLuPop06} given by metric~\eqref{2spins_Chen} with~\eqref{QP_Chen_gen} (this corresponds to the Einstein spacetime~(48,\cite{CheLuPop06}) with vanishing NUT parameters -- i.e., $L_1=0=L_2$ in the notation of \cite{CheLuPop06} -- and two non-zero spins). As proven in \cite{Hamamotoetal07}, these metrics are of type~D.

\subsubsection{Double copy}

\label{subsubsec_DC}

Having observed that the general solution~\eqref{CarterPlebanskiEF} is of the KS form, it is natural to test if it gives rise to an instance of the KS double copy \cite{MonOCoWhi14} as extended to backgrounds of constant curvature \cite{BahLunWhi17,CarPenTro18}. Indeed, it can be easily checked that the electromagnetic field $\bF=\d\bA$ produced by the potential (aligned along the KS direction)
\be
 \bA=\dfrac{er}{(r^2+y_1^2)(r^2+y_2^2)}\left[\mathrm{d}u + \left( y_1^2+y_2^2 \right)\mathrm{d}\phi_1 + y_1^2y_2^2\mathrm{d}\phi_2 \right] ,
 \label{KS_potential}
\ee
where $e$ is a constant, satisfies the sourcefree Maxwell equations both in the full spacetime~\eqref{CarterPlebanskiEF} and in the KS ``background'' given by~\eqref{CarterPlebanskiEF} with $\mu=0$ (cf. \cite{MyePer86,OrtSri24} for related comments).
 
The potential~\eqref{KS_potential} is gauge equivalent to one of the Maxwell fields found earlier in any dimensions in \cite{CheLu08,Krtous07}. As noticed in those references, a more general solution can be obtained by taking a linear combination of~\eqref{KS_potential} with the additional terms $y_1(r^2+y_1^2)^{-1}(y_2^2-y_1^2)^{-1}\left[ \mathrm{d}u + \left( y_2^2-r^2 \right)\mathrm{d}\phi_1 - r^2y_2^2\mathrm{d}\phi_2 \right]$ and $y_2(r^2+y_2^2)^{-1}(y_1^2-y_2^2)^{-1}\left[ \mathrm{d}u + \left( y_1^2-r^2 \right)\mathrm{d}\phi_1 -r^2y_1^2\mathrm{d}\phi_2 \right]$. These Maxwell fields can also be constructed using the Killing 1-form “background subtraction” method of \cite{Aliev07,Aliev07_2} (cf. also \cite{FroKrtKub17}), and have been recently discussed in the context of the (multi-)KS double copy in \cite{ChaKee23}. The second mWAND of spacetime~\eqref{CarterPlebanskiEF} (cf. appendices~\ref{subsubsec_CarterPlebanski} and \ref{subsubsec_generalized}, and section~3.2 of \cite{PraPraOrt07} for related comments) also defines a KS vector field, however, this leads to a potential which is gauge equivalent to~\eqref{KS_potential}.

\subsection{Recovering the doubly-spinning Kerr-(A)dS metric}

\label{subsubsec_recovering}

Physical and geometric properties of the general solution obtained above will depend on the values chosen for the integration constants $\mu$, $\hat{\mathcal{U}}^0$, $c_0$ and $d_0$ which enter the metric via~\eqref{U0_Final} and \eqref{Omegas_Final}. Let us discuss how these are constrained, and what particular choice gives rise to the doubly-spinning Kerr-(A)dS black holes of \cite{Gibbonsetal04}. Here we will {\em not} assume the gauge~\eqref{gauge_EFCP}, as a different choice will prove more convenient for the purposes of comparing with \cite{Gibbonsetal04}.

The functions $(\omega^0_2)^2$ and $(\omega^0_4)^2$ in~\eqref{Omegas_Final} must both be strictly positive (recall~\eqref{b_and_omega_Equivalence}) in a suitable range of $(y_1,y_2)$, which means the numerators in~\eqref{Omegas_Final} must thus have opposite signs (for a given choice of $(y_1,y_2)$). Therefore, $c_0$ and $d_0$ must be such that the polynomial ${\cal P}(s)$ 
possesses at least a real, positive root (from now on it is convenient to think of ${\cal P}$ as a cubic polynomial in the variable $s^2$, hence having three roots). The remaining two roots of ${\cal P}(s)$ are necessarily either both real or both purely imaginary and complex conjugates. Closer inspection of~\eqref{Omegas_Final} reveals that for $\lambda>0$ the three roots must all be real and distinct, while for $\lambda<0$ no further restrictions occur. From now on, let us restrict ourselves to the case
\begin{align} 
\label{Polynomial completely factorized}
       {\cal P}(s)= \lambda (s^2 - s_0)(s^2-s_1)(s^2-s_2) , \qquad s_0,s_1,s_2\in\mathbbm{R} , \quad s_1\neq s_2. 
\end{align}
(We will not discuss in more detail the remaining cases $s_0=s_1=s_2\in\mathbbm{R}^+$ and $s_0\in\mathbbm{R}^+$, $s_2=\bar s_1\in i\mathbbm{R}$, which can both occur only for $\lambda<0$, and are anyway subcases of the solution of \cite{CheLuPop06} mentioned above.)

Comparing~\eqref{Polynomial completely factorized} with \eqref{Omegas_Final} gives  $2\hat{\mathcal{U}}^0=-\lambda(s_0+s_1+s_2)$. Let us now relabel the root $s_0$ as
\be
 2\mathcal{U}^0 \equiv -\lambda s_0 ,
\label{2U0Definition}
\ee
such that $2\hat{\mathcal{U}}^0=2\mathcal{U}^0-\lambda (s_1+s_2)$. Then eq.~\eqref{U0_Final} becomes
\begin{align}
    2U^0 = 2\mathcal{U}^0+\lambda \left( y_1^2+y_2^2-s_1-s_2 \right) .
		\label{U0_gener}
\end{align}

Recalling the comment after \eqref{rescal_e0_f0}, it is now convenient to choose $e_0$ and $f_0$ such that $s^4-e_0s^2-f_0 = (s^2-s_1)(s^2-s_2)$, which reduces~\eqref{xi2uandxi4u_Integrated} to 
\begin{align} \label{xi2u0Omega20Relation}
    (\lambda y_1^2+2\mathcal{U}^0)\xi_2^{u0} = \omega_2^0, \qquad  (\lambda y_2^2+2\mathcal{U}^0)\xi_4^{u0} = \omega_4^0.
\end{align}
Next, we can use the scaling freedom \eqref{rescaling-begin}--\eqref{rescaling-end} with (see \cite{Ortaggio17})
\begin{align}
    s_1^\prime = \dfrac{s_1}{\kappa^2}, \qquad s_2^\prime = \dfrac{s_2}{\kappa^2}, \qquad \alpha^\prime = \kappa^5 \alpha, \qquad \gamma^\prime = \kappa^5 \gamma, \qquad \beta^\prime = \kappa^3 \beta, \qquad \delta^\prime = \kappa^3 \delta,
\end{align}
with $\kappa = \sqrt{2\lvert \mathcal{U}^0 \rvert}$ to set
\begin{align} \label{epsilon_Definition}
     2{\mathcal{U}^0}'=\epsilon ,  \qquad \epsilon \equiv \mathrm{sign} (\mathcal{U}^0)= \pm 1, 0  .
\end{align}
In other words, we have normalized the root $s_0$ to the convenient value $-\epsilon/\lambda$.

Finally, let us choose $\alpha' = (s_1-s_2)^{-1}$, $\beta' = s_1(s_2-s_1)^{-1}$, $\gamma' = (s_2-s_1)^{-1}$, $\delta' = s_2(s_1-s_2)^{-1}$ (recall these constants entered \eqref{xi2andxi3_final} and \eqref{xi4andxi5_final} arbitrarily and are {\em not} integration constants). With this choice and dropping the primes, we can summarize the asymptotic quantities obtained for this branch of solutions (in a parameter range such that \eqref{Polynomial completely factorized} holds) as follows
\begin{align}
     &\left( \omega^0_2 \right)^2 = \dfrac{\left(\epsilon + \lambda y_1^2  \right)\left(y_1^2-s_1\right)\left(y_1^2-s_2\right)}{y_2^2-y_1^2}, \qquad  \left( \omega^0_4 \right)^2 = \dfrac{\left( \epsilon + \lambda y_2^2  \right)\left(y_2^2-s_1\right)\left(y_2^2-s_2\right)}{y_1^2-y_2^2}, \qquad  \omega_3^0 = 0 = \omega_5^0, \label{omegas_gener} \\
     &\xi_2^{u0} = ( \epsilon + \lambda y_1^2 )^{-1}\omega_2^0, \qquad \xi_4^{u0} =  ( \epsilon + \lambda y_2^2 )^{-1}\omega_4^0, \qquad \xi_3^0 = 0 = \xi_5^0, \\
     &\xi_2^{\alpha 0}\partial_\alpha = \dfrac{\omega_2^0}{\epsilon + \lambda y_1^2}\left( \dfrac{1}{y_1^2-s_1}\partial_{\phi_1} + \dfrac{1}{y_1^2-s_2}\partial_{\phi_2}\right), \qquad \xi_3^{\alpha 0 }\partial_\alpha = \omega_2^0 \partial_{y_1}, \\
     &\xi_4^{\alpha 0}\partial_\alpha = \dfrac{\omega_4^0}{\epsilon + \lambda y_2^2}\left( \dfrac{1}{y_2^2-s_1}\partial_{\phi_1} + \dfrac{1}{y_2^2-s_2}\partial_{\phi_2}\right), \qquad \xi_5^{\alpha 0 }\partial_\alpha = \omega_4^0 \partial_{y_2}, \\
     &2\mathcal{U}^0 = \epsilon, \qquad X^{A0} = \delta^A_u. \label{U0X0_gener}
\end{align}
With~\eqref{U0_gener}, this fully specifies the contravariant metric~\eqref{InverseMetricin6D_General}, which upon inversion results in 
\begin{align}
    \mathrm{d}s^2 &= -\left[\epsilon - \lambda(r^2 + s_1 + s_2 -y_1^2 - y_2^2)\right]\mathrm{d}u^2+2\mathrm{d}r\left[\mathrm{d}u + \dfrac{(s_1-y_1^2)(s_1-y_2^2)}{s_1-s_2}\mathrm{d}\phi_1 + \dfrac{(s_2-y_1^2)(s_2-y_2^2)}{s_2-s_1}\mathrm{d}\phi_2 \right] \nonumber \\ 
		& {}+ (r^2+y_1^2)\dfrac{y_2^2-y_1^2}{{\cal P}(y_1)}\mathrm{d}y_1^2 +(r^2+y_2^2)\dfrac{y_1^2-y_2^2}{{\cal P}(y_2)}\mathrm{d}y_2^2 \nonumber \\ &+(r^2+s_1)(\epsilon+\lambda s_1)\dfrac{(s_1-y_1^2)(s_1-y_2^2)}{s_1-s_2}\mathrm{d}\phi_1^2 +(r^2+s_2)(\epsilon+\lambda s_2)\dfrac{(s_2-y_1^2)(s_2-y_2^2)}{s_2-s_1}\mathrm{d}\phi_2^2 \nonumber \\ 
		& {}+ 2\lambda \mathrm{d}u \left[ (r^2+s_1)\dfrac{(s_1-y_1^2)(s_1-y_2^2)}{s_1-s_2}\mathrm{d}\phi_1 + (r^2+s_2)\dfrac{(s_2-y_1^2)(s_2-y_2^2)}{s_2-s_1}\mathrm{d}\phi_2\right] \nonumber \\ 
		& {}+ \dfrac{\mu r}{(r^2+y_1^2)(r^2+y_2^2)}\left[  \mathrm{d}u + \dfrac{(s_1-y_1^2)(s_1-y_2^2)}{s_1-s_2}\mathrm{d}\phi_1 + \dfrac{(s_2-y_1^2)(s_2-y_2^2)}{s_2-s_1}\mathrm{d}\phi_2 \right]^2 ,
	\label{KerrAdS}	
\end{align}
with~\eqref{Polynomial completely factorized}, \eqref{2U0Definition}, \eqref{epsilon_Definition}. This coincides with the line-element~\eqref{FinalMetricForU0neq0andNonEqualRotations} with $s_1=a_1^2$, $s_2=a_2^2$ and $\mathrm{d}\phi_1=\mathrm{d}\varphi_1/\left[a_1(\epsilon + \lambda a_1^2)\right]$, $\mathrm{d}\phi_2=\mathrm{d}\varphi_2/\left[a_2(\epsilon + \lambda a_2^2)\right]$. 
The form~\eqref{KerrAdS} of the metric is singular for the special parameter choice $s_1=-\epsilon/\lambda$ (or $s_2=-\epsilon/\lambda$), which for $\lambda<0$ and $\epsilon=+1$ has been related to an ``ultraspinning'' configuration \cite{HawHunTay99,CalEmpRod08,Caldarellietal12,Gnecchietal14,Klemm14,HenManKub15,Hennigaretal15} (cf. also a comment in appendix~\ref{subsubsec_CarterPlebanski}). Similarly as~\eqref{CarterPlebanskiEF}, also~\eqref{KerrAdS} is manifestly KS, now with KS covector  given by $\ell_a\d x^a=\mathrm{d}u + (s_1-y_1^2)(s_1-y_2^2)(s_1-s_2)^{-1}\mathrm{d}\phi_1 + (s_2-y_1^2)(s_2-y_2^2)(s_2-s_1)^{-1}\mathrm{d}\phi_2$ (note that the coordinates $(u,\phi_1,\phi_2)$ used here are a linear combination of those of section~\ref{subsubsec_general_metric} -- this is due to the different gauges used in both sections).

In the generic case $b_{45}\neq \pm b_{23}$ with $\mathrm{d}b_{23} \neq 0 \neq \mathrm{d}b_{45}$, we have thus obtained the most general vacuum solution such that the metric function~\eqref{Polynomial completely factorized} is factorized. Thanks to the normalization~\eqref{epsilon_Definition}, the final form of the metric is characterized by three continuous parameters $\mu$, $s_1$ and $s_2$, related to mass and angular momenta, and one discrete constant $\epsilon=\pm1,0$. The positivity of $\left( \omega^0_2 \right)^2$ and $\left( \omega^0_4 \right)^2$ in~\eqref{omegas_gener} gives the constraints on $s_1$ and $s_2$ summarized in table~\ref{table_s1s2}. 
\begin{table}[t]
\label{table_s1s2}
\centering
\begin{tabular}{|l||l|l|} 
\hline
              & $\lambda>0$ & $\lambda<0$  \\ 
\hhline{|=::==|}
$\epsilon=+1$  &    $s_1,s_2>0$        &            \\ 
\hline
$\epsilon=0$  &    $s_1,s_2>0$     &         $s_1>0$        \\ 
\hline
$\epsilon=-1$ &    $s_1>0$         &     $s_1>0$          \\
\hline
\end{tabular}
\caption{{\footnotesize Constraints on the signs of $s_1$ and $s_2$ in~\eqref{omegas_gener}--\eqref{U0X0_gener} (up to permutation $s_1\leftrightarrow s_2$). The sign of a root is understood to be arbitrary when no comments are provided.}}
\end{table}

According to the value of $\epsilon$, the following three distinct families of solutions are obtained:

\begin{enumerate}

	\item $\epsilon = +1$: this is locally isometric to the original solution of~\cite{Gibbonsetal04}, which includes \textit{doubly spinning Kerr-(A)dS black holes} (with unequal spins). Killing horizons are defined by the zeros of the function ${\cal Q}(r)$ defined in~\eqref{Q_main} \cite{Gibbonsetal04}. 
		See appendix~\ref{Section: Appendix: Kerr-(A)dS metrics in six dimensions} for more details, including the relation to the coordinates used in~\cite{Gibbonsetal04}, and \cite{KolKrt17} for a discussion of the proper range of the coordinates $y_1$ and $y_2$. .

	\item $\epsilon = -1$: this branch can alternatively be obtained as an analytic continuation \cite{KleMorVan98,Klemm98,Ortaggio17,MarPeo22_b,ChrConGra25} of the case $\epsilon = +1$ (see again appendix~\ref{Section: Appendix: Kerr-(A)dS metrics in six dimensions} for alternative coordinates). In the $\lambda<0$ case, an analysis of possible compactifications (or a lack thereof) of the four-dimensional subspaces of constant $r$ and $u$ can be found in~\cite{ChrConGra25} (see also \cite{KleMorVan98,Klemm98}).
	
	\item $\epsilon =0$: for $\lambda>0$, this was obtained in \cite{MarPeo22_b} as an infinite-rotation limit of the case $\epsilon = +1$ (cf. \cite{MarPaeSen17,MarPaeSen18} in four dimensions and appendix~\ref{Section: Appendix: Kerr-(A)dS metrics in six dimensions} for more details).
	
\end{enumerate}

\section*{Acknowledgments}

Supported by the Institute of Mathematics, Czech Academy of Sciences (RVO 67985840) and research grant GA25-15544S.

\setcounter{equation}{0}

\section*{Appendices}
\appendix
\addcontentsline{toc}{section}{Appendices}

\renewcommand{\thesection}{\Alph{section}}
\setcounter{section}{0}

\renewcommand{\theequation}{{\thesection}\arabic{equation}}
\setcounter{equation}{0}

\section{NP formalism in the case of a non-degenerate mWAND~$\bm{\ell}$}

\label{app_NP}

In this appendix we summarize the notation and the equations of the NP formalism needed in the present paper. For the null frame defined in section~\ref{subsec_assumpt}, the Ricci rotation coefficients $L_{ab}$, $N_{ab}$ and $\overset{i}{M_{ab}}$ are expressed by \cite{Pravdaetal04}
\begin{align} \label{Appendix_RicciRotationCoefficients}
    L_{ab} = \ell_{a;b} \equiv \ell_{\mu ; \nu} m^{\mu}_{(a)} m^{\nu}_{(b)}, \qquad  N_{ab} = n_{a;b} \equiv n_{\mu ; \nu} m^{\mu}_{(a)} m^{\nu}_{(b)}, \qquad  \overset{i}{M_{ab}} = \overset{(i)}{m_{a;b}} \equiv \overset{(i)}{m_{\mu ; \nu}} m^{\mu}_{(a)} m^{\nu}_{(b)},
\end{align}
and satisfy $L_{0a} = N_{1a} = N_{0a} + L_{1a} = \overset{i}{M_{0a}} + L_{ia} = \overset{i}{M_{1a}} + N_{ia} = \overset{i}{M_{ja}} + \overset{j}{M_{ia}} = 0$. The symmetric and antisymmetric parts of the purely spatial components of $L_{ab}$ will be denoted as
\be
 S_{ij}\equiv L_{(ij)} , \qquad A_{ij}\equiv L_{[ij]} ,
\ee
with $S\equiv S_{ii}$.

From now on, the null vector field $\bm{\ell}$ is a geodesic and affinely parametrized mWAND, and we take a null frame parallelly propagated along it (as in the main text) -- i.e., eqs.~\eqref{GeodandAffineCondition} and \eqref{ParaFrame} hold. We can further use a frame freedom to enforce~\eqref{Li1}, i.e., hereafter we will have
\begin{align}
    L_{i1} = 0 .
\end{align}

Covariant derivatives in the direction of the frame vectors $\bm{\ell}$, $\bm{n}$ and $\bm{m}_{(i)}$ are denoted, respectively, 
\begin{align}
    D \equiv \ell^a\nabla_a, \qquad \Delta \equiv n^a\nabla_a, \qquad  \delta_i \equiv m_{(i)}^a\nabla_a.
\end{align}

Regarding the non-zero Weyl scalars, we mostly follow the notation of \cite{Durkeeetal10}. In general we thus have the  purely spatial components $C_{ijkl}$, accompanied by
\beq
   & & \Phi_{ij} \equiv C_{0i1j}, \qquad \Phi^S_{ij} \equiv \Phi_{(ij)}, \qquad \Phi^A_{ij} \equiv \Phi_{[ij]}, \qquad \Phi \equiv \Phi_{ii}, \label{Weyl0} \\
	 & & \mypsi_{ijk} \equiv C_{1ijk}, \qquad \mypsi_{i} \equiv C_{101i}, \qquad \myomega_{ij} \equiv C_{1i1j} , \label{Weyl<0}
\eeq
satisfying $\myomega_{ij} = \myomega_{ji}$, $\myomega_{ii} = 0$, $\mypsi_{i} = \mypsi_{kik}$, $\mypsi_{ijk} = -\mypsi_{ikj}$, $\mypsi_{[ijk]} = 0$, and
\be 
	2\Phi^S_{ij} = -C_{ikjk} . 
	\label{PhiS}
\ee	
The following equations hold for an $n$-dimensional $\Lambda$-vacuum (i.e., Einstein) spacetime, such that $ R_{a b} = \frac{2\Lambda}{n-2}g_{ab} \equiv (n-1)\lambda g_{ab}$.

In an Einstein spacetime, the Bianchi identities $R_{ab[cd;e]}=0$ reduce to $C_{ab[cd;e]} = 0$, which were given in \cite{Pravdaetal04} in the NP formalism (they are thus affected by $\Lambda$ only implicitly through the Ricci rotation coefficients \eqref{N_ij}, \eqref{Ni1} and \eqref{RicciRotCoeff_integrated}). Under the above setup, eqs.~((B.5) and (B.12), \cite{Pravdaetal04}) take the form (see also, e.g., \cite{PraPraOrt07, OrtPraPra09b})
\begin{align}
    D\Phi_{ij} &= -\Phi L_{ij} - \Phi_{ik}L_{kj} - 2\Phi^A_{ik}L_{kj},\label{bw0DiffBianchiPhi} \\
    DC_{ijkm} &= 2\Phi_{k[i}L_{j]m} + 2\Phi_{m[j}L_{i]k}-C_{ijkl}L_{lm}+C_{ijml}L_{lk}-4\Phi^A_{ij}L_{[km]} , 
    \label{bw0DiffBianchiC}
\end{align}
while~(B.1), (B.6), (B.9) and (B.4) of \cite{Pravdaetal04} become
\begin{align}
    D\mypsi_i &= -2\mypsi_kL_{ki} + \delta_i \Phi, \label{B.1}\\
    D\mypsi_{kij} &= -\mypsi_{mij}L_{mk} - \mypsi_iL_{jk} + \mypsi_jL_{ik} -2\delta_k \Phi^A_{ij} - 4\Phi^A_{[i|m}\overset{m}{M_{|j]k}}, \label{B.6}\\
    D\mypsi_{ijk} &= 2\mypsi_{i[k|l}L_{l|j]}+2\delta_{[j}\Phi_{k]i}+2\mypsi_{i}L_{[jk]}+2\Phi_{[k|l}\overset{l}{M}_{i|j]}-2\Phi_{li}\overset{l}{M}_{[jk]}, \label{B.9}\\
    D\myomega_{ij} &= -\myomega_{ik}L_{kj} + \Delta \Phi_{ji} + \delta_j\mypsi_i + \mypsi_iL_{1j} + \Phi N_{ij} - 2\Phi^A_{ik}N_{kj} + \Phi_{ki}N_{kj} + \Phi_{jk}\overset{k}{M}_{i1}+\Phi_{ki}\overset{k}{M}_{j1}+\mypsi_k \overset{k}{M}_{ij}, \label{B.4}
\end{align}
and finally~((B.13), \cite{Pravdaetal04}) reads
\begin{align}
    -\Delta C_{ijkm} + 2\delta_{[k}\mypsi_{m]ij} = \myomega_{im}L_{jk} - \myomega_{jm}L_{ik} + 2\myomega_{[j|k}L_{|i]m} - 4\Phi^A_{ij}N_{[km]} + 2\Phi_{[i|m}N_{|j]k} + 2\Phi_{[j|k}N_{|i]m} + 2\mypsi_{[k|ij}L_{1|m]} \nonumber \\ +2C_{ij[k|l}N_{l|m]} + 2C_{ij[k|l}\overset{l}{M}_{|m]1} + 2C_{[i|lkm}\overset{l}{M}_{|j]1}+ 2\mypsi_{k[i|l}\overset{l}{M}_{|j]m} + 2\mypsi_{m[j|l}\overset{l}{M}_{|i]k} + 2\mypsi_{lij}\overset{l}{M}_{[km]}. \label{B.13}
\end{align}

Next, the commutators \cite{Coleyetal04vsi} reduce to
\begin{align}
    \Delta D - D\Delta &= L_{11}D, \label{B.12}\\
    \delta_i D - D \delta_i &= L_{1i}D + L_{ji}\delta_j, \label{B.13,2}\\
    \delta_i \Delta - \Delta \delta_i &= N_{i1}D-L_{1i}\Delta + (N_{ji} + \overset{j}{M}_{i1})\delta_j, \label{B.14}\\
    \delta_{[i}\delta_{j]} &= N_{[ij]}D + L_{[ij]}\Delta + \overset{k}{M}_{[ij]}\delta_k. \label{B.15}
\end{align}

The Ricci identities (11k), (11i), (11h), (11l), (11o) and (11p) of \cite{OrtPraPra07} become\footnote{The correction of ((11p), \cite{OrtPraPra07}) pointed out in footnote 7 of \cite{OrtPraPra13rev} is included.}
\begin{align}
    \delta_{[j|}L_{i|k]} &= L_{1[j|}L_{i|k]} + L_{il}\overset{l}{M}_{[jk]} + L_{l[j|}\overset{l}{M}_{i|k]}, \label{11k}\\
    \Delta L_{ij} &= L_{11}L_{ij} - L_{kj}\overset{k}{M}_{i1} - L_{ik}\left( N_{kj} + \overset{k}{M}_{j1} \right) - \Phi_{ij} - \lambda \delta_{ij}, \label{11i}\\
    \Delta N_{ij} - \delta_j N_{i1} &= -L_{11}N_{ij} + 2N_{i1}L_{1j} + N_{k1}\overset{k}{M}_{ij} - N_{kj}\overset{k}{M}_{i1} - N_{ik}\left( N_{kj} + \overset{k}{M}_{j1} \right) - \myomega_{ij}, \label{11h}\\
    \delta_{[j|}N_{i|k]} &= -L_{1[j|}N_{i|k]} + N_{i1}L_{[jk]} + N_{il}\overset{l}{M}_{[jk]} + N_{l[j|}\overset{l}{M}_{i|k]}-\dfrac{1}{2}\mypsi_{ijk}, \label{11l}\\
    \Delta \overset{i}{M}_{jk} - \delta_k \overset{i}{M}_{j1} &= N_{j1}L_{ik}-L_{jk}N_{i1} + \overset{i}{M}_{j1}L_{1k} + \overset{i}{M}_{l1}\overset{l}{M}_{jk} - \overset{i}{M}_{lk}\overset{l}{M}_{j1} - \overset{i}{M}_{jl}\left( N_{lk} + \overset{l}{M}_{k1} \right) - \mypsi_{kij}, \label{11o}\\
    \delta_{[k|}\overset{i}{M}_{j|l]} &= N_{i[l|}L_{j|k]} + L_{i[l|}N_{j|k]} + L_{[kl]}\overset{i}{M}_{j1} + \overset{i}{M}_{p[k|}\overset{p}{M}_{j|l]} + \overset{i}{M}_{jp}\overset{p}{M}_{[kl]} -\dfrac{1}{2}C_{ijkl} - \lambda \delta_{i[k}\delta_{l]j}. \label{11p}
\end{align}

Finally, let us now present a set of \textit{constraint} (i.e., algebraic) equations needed to obtain the exact $r$-dependence of the b.w.~0 Weyl scalars in the next appendix. First of all, ((B.15), \cite{Pravdaetal04}) here reduces to \cite{PraPraOrt07,Durkee09}
\begin{align}\label{bw0algebraicBianchi}
    0 = 2\left( \Phi^A_{jk}L_{im} + \Phi^A_{mj}L_{ik} + \Phi^A_{km}L_{ij} + \Phi_{ij}A_{mk} + \Phi_{ik}A_{jm} + \Phi_{im}A_{kj}   \right) + C_{iljk}L_{lm} + C_{ilmj}L_{lk} + C_{ilkm}L_{lj}.
\end{align}
Further, from the symmetric part of \eqref{bw0DiffBianchiPhi} and a contraction of \eqref{bw0DiffBianchiC} it follows \cite{PraPraOrt07, Durkee09}
\begin{align}\label{C.1}
    0 = -S\Phi_{ij} + \Phi L_{ij} + \Phi_{ik}L_{jk} + 4\Phi^A_{jk}A_{ik} + \left( 2\Phi_{ik} - \Phi_{ki} \right) L_{kj} + 2\Phi^A_{jk}L_{ki}+C_{jkil}L_{lk},
\end{align}
and a contraction of \eqref{bw0algebraicBianchi} leads to \cite{PraPraOrt07,Durkee09}
\begin{align}\label{C.2}
    0 = S\Phi^A_{ij} + \Phi A_{ji} - \Phi_{ik}S_{kj} + \Phi_{jk}S_{ki} + 2\left( \Phi^A_{ki}A_{kj} - \Phi^A_{kj}A_{ki} \right) + \dfrac{1}{2}C_{klij}A_{lk}.
\end{align}

In general, another algebraically independent constraint is given by~(14,\cite{TinPra19}). However, it does not provide new information for the class of spacetimes considered in the present paper.

\section{$r$-dependence of $\Phi_{ij}$ in six dimensions}\label{NPeqsPhi}
\setcounter{equation}{0}

\label{app_Phi_6D}

Here we integrate some of the equations presented in appendix~\ref{app_NP} in order to determine the exact $r$-dependence of the Weyl components $\Phi_{ij}$, which is used in section~\ref{subsec_mi}. We confine ourselves to the case of type~II Einstein spacetimes of dimension $n=6$. Conditions~\eqref{GeodandAffineCondition}, \eqref{DetofL}, \eqref{ParaFrame}, \eqref{Li1}, \eqref{bw0WeylAsymBehaviour}, \eqref{l1i_vanishing}, \eqref{l11Vanish} will be understood. In this particular case, without loss of generality one can assume the form~\eqref{Lin6D} of the optical matrix, i.e., $L_{22} = r/(r^2+b_{23}^2)=L_{33}$, $L_{23} = b_{23}/(r^2+b_{23}^2)=-L_{32}$ (the second block can be obtained from the first one by interchanging the labels $23$ for $45$), $L_{24} = L_{25} = L_{34} = L_{35} = 0 =L_{42} = L_{52} = L_{43} = L_{53}$. These conditions  will be important for the following analysis, in particular
\begin{align}
     A_{23} = L_{23}, \quad A_{45} = L_{45} .
\end{align}
It will also be useful to remember the spatial index symmetries of the Weyl tensor, i.e., $C_{i[jkl]} = C_{ij(kl)} = C_{(ij)kl} = 0$ and $C_{ijkl} = C_{klij}$, as well as 
the algebraic relation~\eqref{PhiS}, so that, for instance, $-2\Phi^S_{23} = C_{2434} + C_{2535}$, etc..

First, taking the algebraic condition \eqref{C.1} for $ij=22$, $ij=33$ and $ij=23$ leads to, respectively,
\beqn
    & & 4L_{44}\Phi_{22}=\Phi L_{22}+C_{2323}(L_{22}-L_{44}), \label{algebaricPhi22} \\
    & & 4L_{44}\Phi_{33}=\Phi L_{22}+C_{2323}(L_{22}-L_{44}), \\
	  & & 0=-2L_{44}(\Phi_{23}+\Phi^S_{23})+L_{23}(\Phi +C_{2323})+L_{45}C_{2345}. \label{constraint2}
\eeqn
The first two of these equations imply
\begin{align}\label{Phi22Phi33}
    \Phi_{22} = \Phi_{33}.
\end{align}

Next, from \eqref{C.2} for $ij = 23$ it follows
\begin{align}\label{constraint}
    0 = -2L_{44}\Phi^A_{23}+L_{23}(\Phi +C_{2323})+L_{45}C_{2345},
\end{align}
which compared with~\eqref{constraint2} results in
\begin{align}\label{Phi23Phi32}
    \Phi_{23}=-\Phi_{32}.
\end{align}
Similarly, for the second block ($i,j \in \{4,5\}$) one obtains
\begin{align}\label{SecondBlock}
    \Phi_{44} = \Phi_{55}, \quad \Phi_{45} = -\Phi_{54} .
\end{align}

Under these simplifications, the \enquote{block-diagonal} elements (i.e., those within the $2\times2$ blocks corresponding to the structure of the optical matrix~\eqref{Lin6D}) of the differential equation~\eqref{bw0DiffBianchiPhi} read
\begin{align}\label{BlockDiagonalPhi}
   D\Phi_{22}&= -(\Phi +\Phi_{22})L_{22}+3\Phi_{23} L_{23}
   , \qquad
   D\Phi_{23}= -(\Phi +\Phi_{22})L_{23}-3\Phi_{23} L_{22}
   , \nonumber \\
   D\Phi_{44}&= -(\Phi +\Phi_{44})L_{44}+3 \Phi_{45} L_{45}
  , \qquad
   D\Phi_{45}= -(\Phi +\Phi_{44})L_{45}-3 \Phi_{45} L_{44} .
\end{align}
These equations can be solved explicitly (using~\eqref{Lin6D}), but the full solution is cumbersome and not particularly illuminating at this stage. 
We will thus present it below in \eqref{BlockDiagonalPhiFinal} after further simplifications. For now, let us only note that its asymptotic form reads
\beqn
	 & & \Phi_{22} =\frac{\Phi_0^{(1)}}{r}+\frac{\Phi_0^{(1)}b_{45}^2}{r^3}+\frac{\Phi_0^{(2)}b_{23}+2\Phi_0^{(3)}b_{45}^2}{r^4}+\frac{3(\Phi_0^{(4)}-\Phi_0^{(1)}b_{23}^2b_{45}^2)}{r^5} + O(r^{-6}) , \label{Phi22Integrated} \\
	 & & \Phi_{23} =-\frac{\Phi_0^{(1)}b_{23}}{r^2}+\frac{\Phi_0^{(2)}}{r^3}+\frac{3\Phi_0^{(1)}b_{23}b_{45}^2}{r^4}+2b_{23}\frac{\Phi_0^{(2)}b_{23}+2\Phi_0^{(3)}b_{45}}{r^5} + O(r^{-6}) , \label{Phi23Integrated} 
\eeqn
along with similar expressions for $\Phi_{44}$ and $\Phi_{45}$, where $\Phi_0^{(\aleph)}$ ($\aleph = 1, \dots, 4$) are four integration functions independent of~$r$.

Let us now show how those integration functions can be constrained. To this end, we first need to determine also the $r$-dependence of the Weyl components $C_{2323}$, $C_{2345}$ and  $C_{4545}$. From~\eqref{bw0DiffBianchiC} one has
\begin{align}\label{C2323_C2345}
    DC_{2323}& =2L_{22}(\Phi_{22}-C_{2323})-6\Phi_{23}L_{23}, \qquad DC_{4545}=2L_{44}(\Phi_{44}-C_{4545})-6\Phi_{45}L_{45}  , \nonumber \\ 
		DC_{2345}&=-4\Phi_{23} L_{45}-2 L_{44} C_{2345} , 
\end{align}
which can, again, be integrated exactly. The asymptotic form of the solution is (cf.~\eqref{BlockDiagonalPhiFinal2} for its exact expression after further simplifications)
\be
   C_{2323}=\frac{2\Phi_0^{(1)}}{r} + \dfrac{C^{(0)}_{2323}}{r^2} + O(r^{-3}) , \qquad C_{2345}=\dfrac{C^{(0)}_{2345}}{r^2} + O(r^{-3}) ,  \qquad C_{4545}=-\frac{2\Phi_0^{(1)}}{r} + \dfrac{C^{(0)}_{4545}}{r^2} + O(r^{-3}) ,
\ee
where $C^{(0)}_{2323}$, $C^{(0)}_{2345}$ and $C^{(0)}_{4545}$ are new integration functions.

However, from~\eqref{bw0WeylAsymBehaviour} one gets\footnote{The first of~\eqref{cond_C_2323_C2345} can be obtained also if the assumption~\eqref{bw0WeylAsymBehaviour} is relaxed \cite{OrtPra14}.}
\be
  \Phi_0^{(1)}=0 , \qquad C^{(0)}_{2323}=C^{(0)}_{2345}=C^{(0)}_{4545}=0 . \label{cond_C_2323_C2345}
\ee

Next, eq.~\eqref{constraint} (with its counterpart under interchanging the indices $(23)\leftrightarrow(45)$) results in
\begin{align}
   \Phi_0^{(2)}=0=\Phi_0^{(3)} .
\end{align}
We are thus left with only one arbitrary integration function, which we relabel more compactly as 
\begin{align} \label{Phi_0Definition}
    \Phi_0^{(4)}\equiv\Phi_0 .
\end{align}

The exact $r$-dependence of the ``block-diagonal'' elements of $\Phi_{ij}$ thus finally reads
\begin{align}
    \Phi_{22} &=  \dfrac{\Phi_0r\left[ 3r^4+(b_{45}^2-b_{23}^2)r^2-3b_{23}^2b_{45}^2 \right]}{(r^2+b_{23}^2)^3(r^2+b_{45}^2)^2}  , \qquad \Phi_{23}=\dfrac{\Phi_0b_{23}\left[ b_{45}^2(3r^2-b_{23}^2)+r^2(5r^2+b_{23}^2)) \right]}{(r^2+b_{23}^2)^3(r^2+b_{45}^2)^2}, \label{BlockDiagonalPhiFinal} \\ 
		 C_{2323} &= -\dfrac{2\Phi_0r(r^2-3b_{23}^2)}{(r^2+b_{23}^2)^3(r^2+b_{45}^2)}, \qquad C_{2345} = \dfrac{4\Phi_0b_{23}b_{45}r}{(r^2+b_{23}^2)^2(r^2+b_{45}^2)^2}, \qquad  C_{4545}=-\dfrac{2\Phi_0r(r^2-3b_{45}^2)}{(r^2+b_{45}^2)^3(r^2+b_{23}^2)} , \label{BlockDiagonalPhiFinal2}
\end{align}
along with similar expressions for $\Phi_{44}$ and $\Phi_{45}$, which can be obtained from $\Phi_{22}$ and $\Phi_{23}$ by just interchanging the indices $(23)\leftrightarrow(45)$.

For the remaining components, eq.~\eqref{bw0DiffBianchiPhi} reads
\begin{align}\label{OffBlockDiagonalPhi}
    D\Phi_{42} &= -(2\Phi_{42}-\Phi_{24})L_{22}+(2\Phi_{43}-\Phi_{34})L_{23}, \quad     
    D\Phi_{43} = -(2\Phi_{43}-\Phi_{34})L_{22}-(2\Phi_{42}-\Phi_{24})L_{23} , \nonumber \\
    D\Phi_{52} &= -(2\Phi_{52}-\Phi_{25})L_{22}+(2\Phi_{53}-\Phi_{35})L_{23}, \quad 
     D\Phi_{53} = -(2\Phi_{53}-\Phi_{35})L_{22}-(2\Phi_{52}-\Phi_{25})L_{23} .
\end{align}
Let us now show that the above off-block-diagonal elements must in fact vanish. This has to be done separately for the two possible cases $b_{45}^2-b_{23}^2\neq0$ and $b_{45}^2-b_{23}^2=0$ (cf., e.g., the denominator in eq.~\eqref{C2545} below).\footnote{From section~\ref{sec_generic} of the main text we enforce assumption~\ref{ass2b}. of section~\ref{subsec_summary}, which rules out the case $b_{45}^2-b_{23}^2=0$. Nevertheless, we include it in the present discussion for completeness and future purposes.}

\subsection{Generic case $b_{45}^2-b_{23}^2\neq0$}

\label{Section4.1}

First, one can use~\eqref{C.1} and \eqref{C.2} for $ij=24,25,34,35$ to algebraically determine $\Phi_{24}$, $\Phi_{25}$, $\Phi_{34}$, $\Phi_{35}$, $C_{2545}$, $C_{2454}$, $C_{3545}$ and $C_{3454}$ in terms of $\Phi_{42}$, $\Phi_{52}$, $\Phi_{43}$ and $\Phi_{53}$, resulting in
\beqn
   & & \Phi_{24}= \dfrac{-r(r\Phi_{42}+b_{23}\Phi_{43})+b_{45}(r\Phi_{52}+b_{23}\Phi_{53})}{r^2+b_{45}^2}, \qquad     \Phi_{25} = -\dfrac{r(r\Phi_{52}+b_{45}\Phi_{42})+b_{23}(r\Phi_{53}+b_{45}\Phi_{43})}{r^2+b_{45}^2}, \label{OffBlockDiagonalPhiConstraints} \\  
	& & 	\Phi_{34}= \dfrac{b_{23}(r\Phi_{42}-b_{45}\Phi_{52})+r(-r\Phi_{43}+b_{45}\Phi_{53})}{r^2+b_{45}^2}, \qquad     \Phi_{35} = \dfrac{b_{23}(r\Phi_{52}+b_{45}\Phi_{42})-r(r\Phi_{53}+b_{45}\Phi_{43})}{r^2+b_{45}^2}, \label{OffBlockDiagonalPhiConstraints2}	\\
	& & C_{2545}= \dfrac{r \left(b_{23} \Phi_{43}+r \Phi_{42}\right)-b_{45}
   \left(b_{23} \Phi_{53}+r \Phi_{52}\right)}{b_{45}^2+r^2}+2\frac{b_{23}\left(b_{23}\Phi_{42}+b_{45} \Phi_{53}-r\Phi_{43}\right)+b_{45} r \Phi_{52}}{b_{23}^2-b_{45}^2} , \label{C2545}
\eeqn
together with similar expressions for $C_{2454}$, $C_{3545}$ and $C_{3454}$ (recall also~\eqref{PhiS}), which we skip for brevity.

Using~\eqref{OffBlockDiagonalPhiConstraints} and \eqref{OffBlockDiagonalPhiConstraints2}, eqs.~\eqref{OffBlockDiagonalPhi} can be solved explicitly. For brevity, let us display only the asymptotic behaviour of the resulting expressions, i.e., 
\be
 \Phi_{42}=\frac{b_{23} \hat{\Phi}_0^{(3)}+b_{45} \hat{\Phi}_0^{(2)}}{2 r^3}+ O(r^{-4}) , \qquad \Phi_{43}=\frac{-b_{23} \hat{\Phi}_0^{(1)}+b_{45} \hat{\Phi}_0^{(4)}}{2 r^3}+ O(r^{-4}) ,
\ee
and similar expressions for $\Phi_{52}$ and $\Phi_{53}$, where $\hat{\Phi}_0^{(\aleph)}$ ($\aleph = 1, \dots, 4$) are four arbitrary integration functions. From~\eqref{C2545}
(and the corresponding expressions for $C_{2454}$, $C_{3545}$ and $C_{3454}$ that we have not displayed) one thus concludes
\be
  C_{2545}=\frac{\hat{\Phi}_0^{(1)}}{r^2}+ O(r^{-3}) , \quad C_{2454}=\frac{\hat{\Phi}_0^{(2)}}{r^2}+ O(r^{-3}) , \quad C_{3545}=\frac{\hat{\Phi}_0^{(3)}}{r^2}+ O(r^{-3}) , \quad C_{3454}=\frac{\hat{\Phi}_0^{(4)}}{r^2}+ O(r^{-3}) , \\
\ee
However, thanks to~\eqref{bw0WeylAsymBehaviour} one gets $\hat{\Phi}_0^{(\aleph)}=0$, which implies that $C_{2545}$, $C_{2454}$, $C_{3545}$, $C_{3454}$ and all off-block-diagonal $\Phi_{ij}$ also vanish, as we wanted to prove.

\subsection{Special case $b_{45}^2=b_{23}^2\neq0$}

For definiteness, let us discuss only the case $b_{45}=b_{23}$. When $b_{45}=-b_{23}$, some signs will change in the following intermediate expressions, but not the conclusion.

Because of the assumed degeneracy condition on the eigenvalues of $b$, in this case one needs to slightly modify the analysis of section~\ref{Section4.1}. Let us briefly describe how. Using~\eqref{C.1} and \eqref{C.2} for $ij=24,25,34,35$, instead of~\eqref{OffBlockDiagonalPhiConstraints}--\eqref{C2545} one now obtains the relations
\beqn
  & & \Phi_{(24)}=\Phi_{(25)}=\Phi_{(34)}=\Phi_{(35)}=0 , \qquad \Phi_{53}=-\Phi_{42} , \quad \Phi_{43}=\Phi_{52} , \\
	& & C_{3454}=C_{2545}-4\Phi_{42}+\frac{2r}{b_{23}}\Phi_{52} , \qquad C_{3545}=-C_{2454}+4\Phi_{52}+\frac{2r}{b_{23}}\Phi_{42} .
\eeqn

Similarly as before, one can then find the exact $r$-dependence of $\Phi_{42}$, $\Phi_{52}$, $C_{2454}$ and $C_{2545}$ by solving~\eqref{OffBlockDiagonalPhi} and~\eqref{bw0DiffBianchiC} with $ijkm=2454,2545$. Asymptotically, at the (sub-)leading order this gives 
\beqn
 & & \Phi_{42}=\frac{\hat{\Phi}_0^{(1)}}{r^3}+ O(r^{-4}) , \qquad \Phi_{52}=\frac{\hat{\Phi}_0^{(2)}}{r^3}+ O(r^{-4}) , \\
 & & C_{2545}=\frac{C_{2545}^{(0)}}{r^2}-\frac{\hat{\Phi}_0^{(1)}}{r^3}+ O(r^{-4}) , \quad C_{2454}=\frac{C_{2454}^{(0)}}{r^2}-\frac{\hat{\Phi}_0^{(2)}}{r^3}+ O(r^{-4}) , 
\eeqn
where $\hat{\Phi}_0^{(1)}$, $\hat{\Phi}_0^{(2)}$, $C_{2545}^{(0)}$ and $C_{2454}^{(0)}$ are integration functions (unrelated to those denoted by the same symbols in section~\ref{Section4.1}). As before, eq.~\eqref{bw0WeylAsymBehaviour} implies $C_{2545}^{(0)}=0=C_{2454}^{(0)}$. Finally, using~\eqref{C.2} with $ij=42,52$ gives $\hat{\Phi}_0^{(1)}=0=\hat{\Phi}_0^{(2)}$, such that eventually $\Phi_{42}=0=\Phi_{52}$, i.e., all off-block-diagonal $\Phi_{ij}$ vanish, as we wanted to prove.

\subsection{Summary}

Let us summarize the results obtained above. As it turned out (from \eqref{BlockDiagonalPhiFinal} and the fact that all off-block-diagonal components of $\Phi_{ij}$ vanish), the general form of $\Phi_{ij}$ for a six-dimensional type~II spacetime with a non-degenerate optical matrix $L_{ij}$ and a parallelly transported frame, subject to the assumption $C_{ijkm}=o(r^{-2})$, can be conveniently expressed by defining a scalar function
\begin{align}\label{FunctionH}
    H &= \Phi_0 \dfrac{r}{(r^2+b_{23}^2)(r^2+b_{45}^2)} ,
 \end{align}
where $\Phi_0$ is independent of $r$, as 
\begin{align}\label{r-dependencePhi}
\Phi_{2\mu, 2\mu} &= \Phi_{2\mu+1, 2\mu+1} = -2HL^2_{2\mu, 2\mu+1}-L_{2\mu, 2\mu}DH, \nonumber \\
\Phi_{2\mu, 2\mu+1} &= \Phi^A_{2\mu, 2\mu+1} = -D(HL_{2\mu, 2\mu+1}), \nonumber \\
\Phi_{\alpha \beta} &= -\dfrac{1}{r}\delta_{\alpha \beta}DH \qquad (\mbox{for } p=1)
\end{align}
where $\alpha, \beta=4,5$, $\mu = 1, \dots, p$ and $p\equiv\mbox{rank}(b)=1,2$ (where, for definiteness, for $p=1$ we have assumed $b_{23}\neq0=b_{45}$). All the remaining components of $\Phi_{ij}$ vanish. The form of $\Phi_{ij}$
coincides with the one obtained for Kerr--Schild spacetimes in \cite{OrtPraPra09, MalPra11} (although we have not assumed the metric to be Kerr--Schild). The case $p=0$ (excluded for brevity from the above analysis) corresponds to the known Robinson--Trautman spacetimes and can be obtained by setting $b_{23}=0=b_{45}$ in~\eqref{r-dependencePhi} (with~\eqref{Lin6D} and \eqref{FunctionH}), cf.~\cite{OrtSri24,Ortaggio07,OrtPraPra13}.

One still has a freedom of $r$-independent boosts of $\bm{\ell}$, which can be used to make $\Phi_0$ constant \cite{Kinnersley69,deFGodRea15,Ortaggio17} 
(without affecting \eqref{GeodandAffineCondition}, \eqref{ParaFrame} and \eqref{Li1}, and preserving the form of \eqref{Lin6D}, $\Phi_{ij}$ and $C_{ijkm}$, up to suitably rescaling $b_{23}$ and $b_{45}$).

Let us further observe that the trace of $\Phi_{ij}$ is given simply by $\Phi\equiv\Phi_{ii}=-4H(L_{23}^2+L_{45}^2)-2(L_{22}+L_{44})DH$, from which (with~\eqref{FunctionH}, \eqref{r-dependencePhi}) one readily obtains the explicit $r$-dependence of $L_{11}$  by integration of $DL_{11}=-\Phi+\lambda$ (eq.~(11a,\cite{OrtPraPra07})) and thus also of $U$ from $DU=-L_{11}$ (eq.~(21,\cite{Coleyetal04vsi}) applied on $r$), i.e.,
\be
 U=-\frac{\lambda}{2}r^2+U^0+\Phi_0\dfrac{r}{(r^2+b_{23}^2)(r^2+b_{45}^2)} .
 \label{U}
\ee
The latter equation will be useful in the section~\ref{subsec_mi}.

\section{Kerr-(A)dS metrics}

\setcounter{equation}{0}

\label{Section: Appendix: Kerr-(A)dS metrics in six dimensions}

\subsection{General metric in all even dimensions}

In this appendix, we present a unified form (in various coordinate representations) of the $n$-dimensional ($n\ge4$, even) Einstein spacetimes constructed in \cite{Gibbonsetal04} and their extensions obtained in  \cite{MarPeo22_b,ChrConGra25}. For the spatial coordinates $\mu_i$ (with $i=1,\ldots, \frac{n}{2}-1$), we use the normalization of \cite{MarPeo22_b} (which slightly differs from \cite{Gibbonsetal04,ChrConGra25}).

\subsubsection{Stationary Kerr--Schild coordinates}

Let us start from the following $n$-dimensional line-element in stationary Kerr--Schild coordinates $(\tau, r, \Phi_i, \mu_i, \mu_{n/2})$
\beqn
 \label{KerrSchild metric}
    \mathrm{d}s^2 = -\Xi(\epsilon-\lambda r^2)\mathrm{d}\tau^2 + F\mathrm{d}r^2 + \sum_{i=1}^{n/2-1}\left( r^2+a_i^2 \right)\left(\mathrm{d}\mu_i^2+\mu_i^2 \mathrm{d}\Phi_i^2\right)  + \dfrac{r^2}{\Xi}\mathrm{d}\mu_{n/2}^2\left( \dfrac{\mu_{n/2}^2}{\epsilon - \lambda r^2} +\sum_{i=1}^{n/2-1}\mu_i^2 \right) \nonumber \\ 
		{} + \dfrac{\lambda}{\Xi \left( \epsilon - \lambda r^2 \right)}\left( \sum_{i=1}^{n/2-1} (r^2+a_i^2)\mu_i\mathrm{d}\mu_i \right)\left( 2r^2 \mu_{n/2}\mathrm{d}\mu_{n/2} + \sum_{i=1}^{n/2-1} \epsilon(r^2+a_i^2)\mu_i\mathrm{d}\mu_i \right) \nonumber \\
		{} + \dfrac{\mu r}{\rho^2}\left( \Xi\mathrm{d}\tau + F\mathrm{d}r+\sum_{i=1}^{n/2-1}a_i\mu_i^2\mathrm{d}\Phi_i \right)^2,  
\eeqn
where $a_i$ and $\mu$ are constants, and
\beqn
    & & \Xi \equiv \mu_{n/2}^2 + \sum_{i=1}^{n/2-1}\epsilon \mu_i^2, \qquad F=\dfrac{r^2}{\epsilon - \lambda r^2}\left( \dfrac{\mu_{n/2}^2}{r^2} + \sum_{i=1}^{n/2-1}\dfrac{(\epsilon + \lambda a_i^2)\mu_i^2}{r^2+a_i^2} \right)  ,  \label{F} \\ 
		& & \rho^2 = r^2\left(\dfrac{\mu_{n/2}^2}{r^2}+ \sum_{i=1}^{n/2-1}\dfrac{(\epsilon + \lambda a_i^2)\mu_i^2}{r^2+a_i^2} \right)\prod_{j=1}^{n/2-1}(r^2+a_j^2) ,  \label{rho} \\ 
  & & \mu_{n/2}^2+\sum_{i=1}^{n/2-1}(\epsilon + \lambda a_i^2)\mu_i^2=1 , \qquad \epsilon = \pm 1 , 0 .  \label{constr}
\eeqn

The above metric describes an Einstein spacetime for any value of $\lambda$ and for any $\epsilon = \pm 1 , 0$ (except that it is not defined for $\epsilon=0=\lambda$). In particular, for $\epsilon=+1$ it gives case~(a) in Table~I of \cite{MarPeo22_b}, which (after rescaling $\mu_i^2\mapsto\mu_i^2/(1+\lambda a_i^2)$, $\Phi_i\mapsto-\Phi_i$) corresponds to the original metric of~\cite{Gibbonsetal04} (cf. eqs.~(2.1), (2.3), (2.9)--(2.12) therein). The choice $\epsilon=0$ corresponds to case~(b) in Table~I of \cite{MarPeo22_b} (up to using $\mu_{n/2}\mathrm{d}\mu_{n/2}+\sum_i \lambda a_i^2 \mu_i \mathrm{d}\mu_i=0$, which follows by differentiation of~\eqref{constr}), which for $\lambda>0$ has been interpreted in the sense of an infinite-rotation limit \cite{MarPeo22_b} (cf. also \cite{MarPaeSen17,MarPaeSen18} for earlier results in four dimensions). The branch $\epsilon=-1$ corresponds to case~(c.2) in Table~I of \cite{MarPeo22_b}, which for $\lambda<0$ is also equivalent to~(3.29,\cite{ChrConGra25}) (again up to some coordinate rescalings). At least for some of the signs combinations of table~\ref{table_KerrAdS}, the existence of spacetime regions where~\eqref{KerrSchild metric} admits a Lorentzian signature has been discussed in~\cite{MarPeo22_b,ChrConGra25} (for $n=6$, all $\lambda\neq0$ cases are analyzed in section~\ref{subsubsec_recovering} using different coordinates). The KS covector (which is a geodesic mWAND \cite{OrtPraPra09,MalPra11}) reads $\ell_a\d x^a=\Xi\mathrm{d}\tau + F\mathrm{d}r+\sum_{i=1}^{n/2-1}a_i\mu_i^2\mathrm{d}\Phi_i$.

\begin{table}[t]
\centering
\begin{tabular}{|l||l|l|l|} 
\hline
              & $\lambda>0$ & $\lambda<0$ & $\lambda=0$ \\ 
\hhline{|=::===|}
$\epsilon=+1$  &   \cite{Gibbonsetal04}   &  \cite{Gibbonsetal04}     &   \cite{MyePer86}   \\ 
\hline
$\epsilon=0$  &    \cite{MarPeo22_b}   &     \cite{MarPeo22_b}     &    $\times$    \\ 
\hline
$\epsilon=-1$ &    \cite{MarPeo22_b}    &     \cite{MarPeo22_b,ChrConGra25}     &   \cite{Ortaggio17} ($n=6$) \\
\hline
\end{tabular}
\caption{{\footnotesize The extended even dimensional Kerr-(A)dS metric~\eqref{KerrSchild metric} admits various signs of $\epsilon$ and $\lambda$, as discussed in the references given in the table (refs. \cite{MyePer86,Gibbonsetal04,MarPeo22_b} include also the case when $n$ is odd, while \cite{Ortaggio17} is restricted to the case $n=6$; although the analysis of \cite{MarPeo22_b} focuses on the $\lambda>0$ case, the metrics given there admit also $\lambda<0$). The cases $\epsilon=+1$ and $\epsilon=-1$ are related by analytic continuation \cite{Ortaggio17,MarPeo22_b,ChrConGra25}, while the metric is not defined for $\epsilon=0=\lambda$. When all spin parameters are unequal and non-zero, all cases are also contained (using different coordinates) in the class of Einstein spacetimes of \cite{CheLuPop06}. Special cases of the same metrics (again in different coordinates) appeared also in \cite{Klemm98,HawHunTay99,deFGodRea15} (for the case $n=4$, in particular, cf. \cite{Kerr63,Carter68pla,Carter68cmp,Kinnersley69,Debever71,Carter73,Plebanski75_2,PleDem76,GarciaD84,DebKamMcL84,KleMorVan98,MarPaeSen17,MarPaeSen18}, and \cite{Stephanibook,GriPodbook} for more references).}}
\label{table_KerrAdS}
\end{table}

\subsubsection{Boyer--Lindquist coordinates}

One can recast~\eqref{KerrSchild metric} in Boyer--Lindquist form by defining (similarly as in~\cite{Gibbonsetal04}) 
\begin{align}
\label{Transformation to BL}
    \mathrm{d}\tau = -\mathrm{d}t + \dfrac{\mu r}{\left(\epsilon - \lambda r^2\right)\left(\dfrac{\rho^2}{F}-\mu r\right)}\mathrm{d}r, \qquad \mathrm{d}\Phi_i = \mathrm{d}\psi_i - \lambda a_i\mathrm{d}t - \dfrac{a_i\mu r}{\left( r^2+a_i^2 \right)\left( \dfrac{\rho^2}{F}-\mu r \right)}\mathrm{d}r ,
\end{align}
which results in
\beqn
     \mathrm{d}s^2 = \left( -\epsilon + \lambda r^2+\lambda \sum_{i=1}^{n/2-1}(\epsilon + \lambda a_i^2)a_i^2\mu_i^2 \right)\mathrm{d}t^2+ \dfrac{\rho^2}{\dfrac{\rho^2}{F}-\mu r}\mathrm{d}r^2 + \sum_{i=1}^{n/2-1}\left( r^2+a_i^2 \right)\left(\mathrm{d}\mu_i^2+\mu_i^2 \mathrm{d}\psi_i^2\right) \nonumber \\ 
		{}+ \dfrac{r^2}{\Xi}\mathrm{d}\mu_{n/2}^2\left( \dfrac{\mu_{n/2}^2}{\epsilon - \lambda r^2} +\sum_{i=1}^{n/2-1}\mu_i^2 \right)-2\lambda\mathrm{d}t\sum_{i=1}^{n/2-1}(r^2+a_i^2)a_i\mu_i^2\mathrm{d}\psi_i \nonumber \\
		{}+ \dfrac{\lambda}{\Xi \left( \epsilon - \lambda r^2 \right)}\left( \sum_{i=1}^{n/2-1} (r^2+a_i^2)\mu_i\mathrm{d}\mu_i \right)\left( 2r^2 \mu_{n/2}\mathrm{d}\mu_{n/2} + \sum_{i=1}^{n/2-1} \epsilon(r^2+a_i^2)\mu_i\mathrm{d}\mu_i \right) + \dfrac{\mu r}{\rho^2}\left( -\mathrm{d}t + \sum_{i=1}^{n/2-1}a_i\mu_i^2\mathrm{d}\psi_i \right)^2.
		\label{BL_general}
\eeqn
(Note that $\rho^2 F^{-1}=(\epsilon - \lambda r^2)\prod_{i=1}^{n/2-1}(r^2+a_i^2)$ is a (polynomial) function of $r$ only.) Zeros of $g_{rr}^{-1}$ define Killing horizons \cite{Gibbonsetal04,ChrConGra25}. For $\epsilon = +1$, eq.~\eqref{BL_general} is equivalent to metric~(3.5,\cite{Gibbonsetal04}), while for $\epsilon =-1$ to~(3.5,\cite{ChrConGra25}) (up to redefining $\d\psi_i=\d\bar\psi_i+\lambda a_i\d t$). In these coordinates one has $\ell_a\d x^a= -\mathrm{d}t + \sum_{i=1}^{n/2-1}a_i\mu_i^2\mathrm{d}\psi_i+\rho^2(\rho^2 F^{-1}-\mu r)^{-1}\d r$, with the second mWAND given by $\ell_a\d x^a= -\mathrm{d}t + \sum_{i=1}^{n/2-1}a_i\mu_i^2\mathrm{d}\psi_i-\rho^2(\rho^2 F^{-1}-\mu r)^{-1}\d r$.

Let us briefly comment on the relation of the above parametrization to the standard coordinates used for the familiar $n=4$ Kerr-(A)dS metrics (cf. section~\ref{app_6D_coords} below for the case $n=6$ relevant to the present paper). Let us relabel $a_1\mapsto a$, since here it is the only rotation parameter. If $\epsilon+\lambda a^2\neq0$, by defining $y^2\equiv a^2\mu_2^2=a^2-a^2(\epsilon+\lambda a^2)\mu_1^2$ and shifting $\d t\mapsto \d t+a(\epsilon+\lambda a^2)^{-1}\d\psi_1$, one obtains metric~(4,\cite{Carter68cmp}) (up to a constant rescaling of $\psi_1$). In the special case $\epsilon=0<\lambda$, metric~(4.24,\cite{MarPaeSen17}) is recovered by further setting $a=1$ and $y=\cos\theta$ (with no need of shifting $t$) -- however, also the choice $\epsilon=0>\lambda$  admits a Lorentzian signature, provided the range of $y$ is restricted to $y^2>a^2$. If $\epsilon+\lambda a^2=0$ (such that $\d\mu_2=0$, cf.~\eqref{constr}) the definition $y\equiv-\frac{1}{2}\mu_1^2$ gives rise to metric~(8,\cite{Carter68cmp}) (with the parameters of \cite{Carter68cmp} set as $e=0=p$, $q=-1$, $h=4\lambda a^2$, $\Lambda=-3\lambda$). This is the unique Taub-NUT metric which belongs to the KS class \cite{OrtSri24} (eq.~(C34,\cite{OrtSri24}) is given in coordinates similar to~\eqref{GeneralMetric}; cf. also (4.44,\cite{MarPaeSen17}) for a different characterization of this spacetime in the case $\lambda>0$).

\subsubsection{Eddington--Finkelstein-like coordinates}

For the purposes of the present paper, it will be convenient to define yet another coordinate transformation 
\begin{align}
    \mathrm{d}\tau=\mathrm{d}u-\dfrac{1}{\epsilon-\lambda r^2}\mathrm{d}r, \qquad \mathrm{d}\Phi_i = \mathrm{d}\varphi_i + \lambda a_i\mathrm{d}u+\dfrac{a_i}{r^2+a_i^2}\mathrm{d}r ,
	\label{Kerr Schild Transformation}
\end{align}
which brings~\eqref{KerrSchild metric} into the Eddington--Finkelstein-like form 
\begin{align} 
\label{GeneralMetric}
    \mathrm{d}s^2 = \left( -\epsilon + \lambda r^2+\lambda \sum_{i=1}^{n/2-1}(\epsilon + \lambda a_i^2)a_i^2\mu_i^2 \right)\mathrm{d}u^2+2\mathrm{d}r\left( \mathrm{d}u + \sum_{i=1}^{n/2-1}a_i\mu_i^2\mathrm{d}\varphi_i \right) + \sum_{i=1}^{n/2-1}\left( r^2+a_i^2 \right)\left(\mathrm{d}\mu_i^2+ \mu_i^2 \mathrm{d}\varphi_i^2\right)\nonumber \\ 
		{}+ \dfrac{r^2}{\Xi}\mathrm{d}\mu_{n/2}^2\left( \dfrac{\mu_{n/2}^2}{\epsilon - \lambda r^2} +\sum_{i=1}^{n/2-1}\mu_i^2 \right) + 2\lambda\mathrm{d}u\sum_{i=1}^{n/2-1}(r^2+a_i^2)a_i\mu_i^2\mathrm{d}\varphi_i \nonumber \\ 
		{}+ \dfrac{\lambda}{\Xi \left( \epsilon - \lambda r^2 \right)}\left( \sum_{i=1}^{n/2-1} (r^2+a_i^2)\mu_i\mathrm{d}\mu_i \right)\left( 2r^2 \mu_{n/2}\mathrm{d}\mu_{n/2} + \sum_{i=1}^{n/2-1} \epsilon(r^2+a_i^2)\mu_i\mathrm{d}\mu_i \right) + \dfrac{\mu r}{\rho^2}\left(\mathrm{d}u + \sum_{i=1}^{n/2-1}a_i\mu_i^2\mathrm{d}\varphi_i \right)^2 . 
\end{align}

Similarly as~\eqref{KerrSchild metric}, also this form of the line-element is manifestly Kerr--Schild, with KS covector $\ell_a\d x^a=\mathrm{d}u + \sum_{i=1}^{n/2-1}a_i\mu_i^2\mathrm{d}\varphi_i$ and $\bl=\pa_r$ (while $\pa_r$ was clearly not a null vector field in the coordinates~\eqref{KerrSchild metric} and \eqref{BL_general}). The asymptotic $(n-1)$-dimensional metric at $r\to\infty$ is conformally flat \cite{MarPeo21} (cf. also \cite{MarPeo22_b,ChrConGra25}.

For $\epsilon=+1$, the metric of $(n-2)$-spaces of constant $r$ and $u$ describes deformed spheres \cite{Gibbonsetal04} (cf.~\cite{MyePer86} in the limit $\lambda=0$). See \cite{KleMorVan98,Klemm98,ChrConGra25} for an analysis of possible compactifications (or a lack thereof) in the remaining cases $\epsilon=0,-1$. For $r\to\infty$, these $(n-2)$-spaces become conformally flat, but not of constant curvature (unless $\lambda=0$ \cite{Ortaggio17}).

In the static limit with all $a_i=0$, metric~\eqref{GeneralMetric} represents Schwarzschild-Tangherlini black holes \cite{Tangherlini63} in Robinson-Trautman coordinates \cite{PodOrt06} (after noticing that $\mu_{n/2} \myd \mu_{n/2}+\epsilon\sum_{i=1}^{n/2-1}\mu_i \myd \mu_i=0$, cf.~\eqref{constr}). In this case, any $(n-2)$-space of constant $u$ and $r$ is of constant curvature, and the parameter $\epsilon = \pm 1 , 0$ determines the sign of its Ricci scalar.

\subsection{Six dimensional metrics with two unequal spins: ``intrinsic'' coordinates and generalized Kerr-(A)dS solution $(a_1^2-a_2^2\neq 0 \neq a_1a_2)$}

\label{app_6D_coords}

\subsubsection{Boyer--Lindquist and Eddington--Finkelstein-like coordinates}

When the two spin parameters are distinct and both non-zero, and if both satisfy $\epsilon+\lambda a_i^2\neq0$, one can conveniently replace the three coordinates $(\mu_1, \mu_2, \mu_3)$ with the constraint~\eqref{constr} by two intrinsic (unconstrained) coordinates $(y_1, y_2)$ defined by \cite{CheLuPop06}
\begin{align}
    \label{mu1andmu2Definition}
    \mu_1^2 \equiv  \dfrac{(a_1^2-y_1^2)(a_1^2-y_2^2)}{a_1^2(a^2_1-a^2_2)(\epsilon+\lambda a_1^2)}, \qquad \mu_2^2 \equiv \dfrac{(a^2_2-y_1^2)(a^2_2-y_2^2)}{a^2_2(a^2_2-a^2_1)(\epsilon+\lambda a_2^2)} , 
\end{align}
such that (cf.~\eqref{rho})
\be
 \rho^2=(r^2+y_1^2)(r^2+y_2^2) .
\ee

The Boyer--Lindquist line-element~\eqref{BL_general} thus becomes 
\beqn
\label{Two Unequal Spins Boyer-Lindquist}
    & & \mathrm{d}s^2 = -\left[\epsilon - \lambda(r^2 + a^2_1 + a^2_2 -y_1^2 - y_2^2)\right]\mathrm{d}t^2 + \dfrac{\rho^2}{(\epsilon-\lambda r^2)(r^2+a_1^2)(r^2+a_2^2)-\mu r}\mathrm{d}r^2 \nonumber \\
		 & & \qquad {} + (r^2+y_1^2)\frac{y_2^2-y_1^2}{{\cal P}(y_1)}\mathrm{d}y_1^2 +(r^2+y_2^2)\frac{y_1^2-y_2^2}{{\cal P}(y_2)}\mathrm{d}y_2^2+ (r^2 + a^2_1)\mu_1^2 \mathrm{d}\psi_1^2 + (r^2 + a^2_2)\mu_2^2 \mathrm{d}\psi_2^2 \nonumber \\ 
		 & & \qquad {} -2\lambda\mathrm{d}t\left[(r^2+a^2_1)a_1\mu_1^2 \mathrm{d}\psi_1 + (r^2+a^2_2)a_2\mu_2^2 \mathrm{d}\psi_2\right]+ 
						\dfrac{\mu r}{\rho^2}(-\mathrm{d}t + a_1\mu_1^2 \mathrm{d}\psi_1+ a_2\mu_2^2 \mathrm{d}\psi_2)^2 ,
\eeqn
while its Eddington--Finkelstein counterpart~\eqref{GeneralMetric} gives rise to 
\beqn 
\label{FinalMetricForU0neq0andNonEqualRotations}
    & & \mathrm{d}s^2 = -\left[\epsilon - \lambda(r^2 + a^2_1 + a^2_2 -y_1^2 - y_2^2)\right]\mathrm{d}u^2 + 2\mathrm{d}r(\mathrm{d}u + a_1\mu_1^2 \mathrm{d}\varphi_1+ a_2\mu_2^2 \mathrm{d}\varphi_2) \nonumber \\
		& & \qquad {} + (r^2+y_1^2)\frac{y_2^2-y_1^2}{{\cal P}(y_1)}\mathrm{d}y_1^2 +(r^2+y_2^2)\frac{y_1^2-y_2^2}{{\cal P}(y_2)}\mathrm{d}y_2^2+ (r^2 + a^2_1)\mu_1^2 \mathrm{d}\varphi_1^2 + (r^2 + a^2_2)\mu_2^2 \mathrm{d}\varphi_2^2 \nonumber \\ 
		& & \qquad {} +2\lambda\mathrm{d}u\left[(r^2+a^2_1) a_1\mu_1^2 \mathrm{d}\varphi_1 + (r^2+a^2_2) a_2\mu_2^2 \mathrm{d}\varphi_2\right] + \dfrac{\mu r}{\rho^2}(\mathrm{d}u + a_1\mu_1^2 \mathrm{d}\varphi_1+ a_2\mu_2^2 \mathrm{d}\varphi_2)^2 ,
\eeqn
where
\be
 {\cal P}(s)\equiv (\epsilon+\lambda s^2)(a^2_1-s^2)(a^2_2-s^2) .
 \label{P_Chen}
\ee

\subsubsection{Carter--Pleba\'nski-like coordinates}

\label{subsubsec_CarterPlebanski}

To conclude, it is useful to present also the Carter--Pleba\'nski-like \cite{Carter68cmp,Plebanski75_2} coordinate representation constructed in \cite{CheLuPop06}. Namely, a linear redefinition of the Killing coordinates 
\beqn
 \d t=-\d\tilde t-(a_1^2+a_2^2)\d\tilde\psi_1-a_1^2a_2^2\d\tilde\psi_2 , \quad \d\psi_1=-a_1(\epsilon + \lambda a_1^2)(\d\tilde\psi_1+a_2^2\d\tilde\psi_2) , \quad \d\psi_2=-a_2(\epsilon + \lambda a_2^2)(\d\tilde\psi_1+a_1^2\d\tilde\psi_2) ,
\eeqn
enables one to rewrite~\eqref{Two Unequal Spins Boyer-Lindquist} as  
\beqn
 \d s^2= (r^2+y_1^2)\frac{y_2^2-y_1^2}{{\cal P}(y_1)}\mathrm{d}y_1^2 +(r^2+y_2^2)\frac{y_1^2-y_2^2}{{\cal P}(y_2)}\mathrm{d}y_2^2+\frac{\rho^2\d r^2}{-{\cal Q}(r)}+\frac{{\cal Q}(r)}{(r^2+y^2_1)(r^2+y^2_2)}\left[\d\tilde t+(y_1^2+y_2^2)\d\tilde\psi_1+y_1^2y_2^2\d\tilde\psi_2\right]^2 \nonumber \\
	  {}+\frac{{\cal P}(y_1)}{(r^2+y^2_1)(y^2_2-y_1^2)}\left[\d\tilde t+(y_2^2-r^2)\d\tilde\psi_1-r^2y_2^2\d\tilde\psi_2\right]^2+\frac{{\cal P}(y_2)}{(r^2+y^2_2)(y^2_1-y_2^2)}\left[\d\tilde t+(y_1^2-r^2)\d\tilde\psi_1-r^2y_1^2\d\tilde\psi_2\right]^2   ,
	\label{2spins_Chen}
\eeqn
with
\be
 {\cal Q}(r)\equiv -(\epsilon-\lambda r^2)(r^2+a_1^2)(r^2+a_2^2)+\mu r .
 \label{Q_Chen}
\ee
(Note that ${\cal Q}=-\rho^2 F^{-1}+\mu r$ in terms of the functions defined in~\eqref{F}, \eqref{rho} with $n=6$.) In these coordinates, the KS covector reads $\ell_a\d x^a=\d\tilde t +(y_1^2+y_2^2)\d\tilde\psi_1+y_1^2y_2^2\d\tilde\psi_2-\rho^2{\cal Q}^{-1}\d r$ and the second mWAND is $\ell_a\d x^a=\d\tilde t +(y_1^2+y_2^2)\d\tilde\psi_1+y_1^2y_2^2\d\tilde\psi_2+\rho^2{\cal Q}^{-1}\d r$ .

For the special value of the spin parameter $a_1^2=-\epsilon/\lambda$, two roots of ${\cal P}(s)$ coincide (cf. \cite{Gnecchietal14,Klemm14} in four dimensions), giving rise to the ``ultraspinning'' (or superentropic) solutions studied in \cite{Hennigaretal15}.\footnote{Ref.~\cite{Hennigaretal15} took the limit $a_1^2\to-1/\lambda$ (in the case $\epsilon=+1$, $\lambda<0$) in coordinates similar to~\eqref{BL_general}. For solutions with a single spin, similar ultraspinning configurations were studied earlier in \cite{HawHunTay99,Caldarellietal12,Gnecchietal14,Klemm14,HenManKub15}.}

\subsubsection{Generalized doubly-spinning Kerr-(A)dS metric}

\label{subsubsec_generalized}

The Einstein space~\eqref{2spins_Chen} can be generalized by replacing ${\cal Q}(r)$ and ${\cal P}(s)$ in~\eqref{P_Chen}, \eqref{Q_Chen} by more general polynomials of the form 
\be
 {\cal Q}(r)=\lambda r^6 -2\hat{\mathcal{U}}^0r^4 - c_0r^2 + \mu r + d_0 , \qquad {\cal P}(s)=\lambda s^6 + 2\hat{\mathcal{U}}^0s^4 - c_0s^2 - d_0 ,
 \label{QP_Chen_gen}
\ee
where $\hat{\mathcal{U}}^0$, $c_0$ and $d_0$ are arbitrary constants, cf.~(48,\cite{CheLuPop06}).\footnote{An additional term linear in $s$ can be added to ${\cal P}(s)$ if one relaxes~\eqref{bw0WeylAsymBehaviour}, which dresses the spacetime with two NUT parameters \cite{CheLuPop06}. There appear to be a few typos in the three equations following~(48,\cite{CheLuPop06}), i.e., the functions $X$, $Y$ and $Z$ should contain $g^2$ instead of $g^6$, and $Z$ should contain $L_2$ instead of $L_1$.} The modulus of any of these can be fixed at will thanks to a scaling freedom\cite{CheLuPop06}, thus reducing the number of essential parameters to three. While the generalization~\eqref{QP_Chen_gen} still gives rise to an Einstein spacetime, it may affect its properties by modifying the root structure of the metric functions. Note, however, that this generalized (i.e., non-factorized) solution concerns only the case $\lambda<0$, so long as one insists on having a Lorentzian signature (cf. section~\ref{subsubsec_recovering} for more details).

Let us finally observe that the coordinate transformation 
\be
     \mathrm{d}\Tilde{t} = \mathrm{d}\bar u + \dfrac{r^4}{{\cal Q}(r)}\mathrm{d}r, \qquad \mathrm{d}\Tilde{\psi}_1 = \mathrm{d}\phi_1 + \dfrac{r^2}{{\cal Q}(r)}\mathrm{d}r, \qquad \mathrm{d}\Tilde{\psi}_2 = \mathrm{d}\phi_2 + \dfrac{1}{{\cal Q}(r)}\mathrm{d}r ,
 \label{transf_CP_EF}		
\ee
brings the line-element~\eqref{2spins_Chen} with~\eqref{QP_Chen_gen} into the Eddington--Finkelstein-like form
\beqn
\label{CarterPlebanskiEF_app}
    & & \mathrm{d}s^2 = 2\mathrm{d}r\left[ \mathrm{d}\bar u + \left( y_1^2+y_2^2 \right)\mathrm{d}\phi_1 + y_1^2y_2^2\mathrm{d}\phi_2 \right] + \left( r^2 + y_1^2 \right)\dfrac{y_2^2-y_1^2}{\mathcal{P}(y_1)}\mathrm{d}y_1^2 + \left( r^2 + y_2^2 \right)\dfrac{y_1^2-y_2^2}{\mathcal{P}(y_2)}\mathrm{d}y_2^2 \nonumber \\ 
		& & \qquad {}+ \dfrac{\mathcal{P}(y_1)}{(r^2+y_1^2)(y_2^2-y_1^2)}\left[ \mathrm{d}\bar u + \left( y_2^2-r^2 \right)\mathrm{d}\phi_1 - r^2y_2^2\mathrm{d}\phi_2 \right]^2+ 
		\dfrac{\mathcal{P}(y_2)}{(r^2+y_2^2)(y_1^2-y_2^2)}\left[ \mathrm{d}\bar u + \left( y_1^2-r^2 \right)\mathrm{d}\phi_1 -r^2y_1^2\mathrm{d}\phi_2 \right]^2 \nonumber \\ 
		& & \qquad {}+\dfrac{\mathcal{Q}(r)}{(r^2+y_1^2)(r^2+y_2^2)}\left[\mathrm{d}\bar u + \left( y_1^2+y_2^2 \right)\mathrm{d}\phi_1 + y_1^2y_2^2\mathrm{d}\phi_2 \right]^2 ,
\eeqn
and transforms the KS covector to $\ell_a\d x^a=\d\bar u +(y_1^2+y_2^2)\d\phi_1+y_1^2y_2^2\d\phi_2$. This coincides with metric~\eqref{CarterPlebanskiEF} obtained in the main text.\footnote{In the present appendix, we have used a barred notation for the coordinate $\bar u$ in order to distinguish it from the coordinate $u$ used in the different Eddington--Finkelstein-like line-element~\eqref{FinalMetricForU0neq0andNonEqualRotations} (and \eqref{GeneralMetric}). When ${\cal Q}$ is given by~\eqref{Q_Chen}, line-elements~\eqref{FinalMetricForU0neq0andNonEqualRotations} and \eqref{CarterPlebanskiEF_app} are related by the transformation $\d u=\d\bar u+(a_1^2+a_2^2)\d\phi_1+a_1^2a_2^2\d\phi_2$, $\d\varphi_1=-a_1(\epsilon + \lambda a_1^2)(\d\phi_1+a_2^2\d\phi_2)$, $\d\varphi_2=-a_2(\epsilon + \lambda a_2^2)(\d\phi_1+a_1^2\d\phi_2)$.
}

%\bibliographystyle{unsrt}
%\bibliography{bibl}

\begin{thebibliography}{10}

\bibitem{Stephanibook}
H.~Stephani, D.~Kramer, M.~MacCallum, C.~Hoenselaers, and E.~Herlt.
\newblock {\em Exact Solutions of {E}instein's Field Equations}.
\newblock Cambridge University Press, Cambridge, second edition, 2003.

\bibitem{GriPodbook}
J.~B. Griffiths and J.~Podolsk\'y.
\newblock {\em Exact Space-Times in {E}instein's General Relativity}.
\newblock Cambridge University Press, Cambridge, 2009.

\bibitem{Carter68pla}
B.~Carter.
\newblock A new family of {E}instein spaces.
\newblock {\em Phys. Lett. {\rm A}}, 26:399--400, 1968.

\bibitem{Carter68cmp}
B.~Carter.
\newblock {H}amilton-{J}acobi and {S}chrodinger separable solutions of
  {E}instein's equations.
\newblock {\em Commun. Math. Phys.}, 10:280--310, 1968.

\bibitem{Kinnersley69}
W.~Kinnersley.
\newblock Type {D} vacuum metrics.
\newblock {\em J. Math. Phys.}, 10:1195--1203, 1969.

\bibitem{Debever71}
R.~Debever.
\newblock On type~{D} expanding solutions of {E}instein-{M}axwell equations.
\newblock {\em Bull. Soc. Math. Belg.}, 23:360--376, 1971.

\bibitem{Carter73}
B.~Carter.
\newblock Black hole equilibrium states.
\newblock In C.~De~Witt and B.~S. De~Witt, editors, {\em Black holes}, pages
  57--214. Gordon and Breach, New York, 1973.

\bibitem{Plebanski75_2}
J.~F. Pleba\'nski.
\newblock A class of solutions of {E}instein-{M}axwell equations.
\newblock {\em Ann. Physics}, 90:196--255, 1975.

\bibitem{PleDem76}
J.~F. Pleba\'nski and M.~Demia\'nski.
\newblock Rotating, charged, and uniformly accelerating mass in general
  relativity.
\newblock {\em Ann. Physics}, 98:98--127, 1976.

\bibitem{GarciaD84}
A.~Garc\'{\i}a~D\'{\i}az.
\newblock Electrovac type~{D} solutions with cosmological constant.
\newblock {\em J. Math. Phys.}, 25:1951--1954, 1984.

\bibitem{DebKamMcL84}
R.~Debever, N.~Kamran, and R.~G. McLenaghan.
\newblock Exhaustive integration and a single expression for the general
  solution of the type~{D} vacuum and electrovac field equations with
  cosmological constant for a nonsingular aligned {M}axwell field.
\newblock {\em J. Math. Phys.}, 25:1955--1972, 1984.

\bibitem{KerSch65}
R.~P. Kerr and A.~Schild.
\newblock A new class of vacuum solutions of the Einstein field equations.
\newblock In G.~Barb\`era, editor, {\em Atti del Convegno sulla Relativit\`a
  Generale: Problemi dell'Energia e Onde Gravitazionali}, pages 1--12, Firenze,
  1965.

\bibitem{KerSch652}
R.~P. Kerr and A.~Schild.
\newblock Some algebraically degenerate solutions of {E}instein's gravitational
  field equations.
\newblock {\em Proc. Symp. Appl. Math.}, 17:199--209, 1965.

\bibitem{DebKerSch69}
G.~C. Debney, R.~P. Kerr, and A.~Schild.
\newblock Solutions of the {E}instein and {E}instein-{M}axwell equations.
\newblock {\em J. Math. Phys.}, 10:1842--1854, 1969.

\bibitem{Tangherlini63}
F.~R. Tangherlini.
\newblock Schwarzschild field in $n$ dimensions and the dimensionality of space
  problem.
\newblock {\em Il Nuovo Cimento}, 27:636--651, 1963.

\bibitem{MyePer86}
R.~C. Myers and M.~J. Perry.
\newblock Black holes in higher dimensional space-times.
\newblock {\em Ann. Phys. (N.Y.)}, 172:304--347, 1986.

\bibitem{Chakrabarti86}
A.~Chakrabarti.
\newblock {K}err metric in eight dimensions.
\newblock {\em Phys. Lett. {\rm B}}, 172:175--179, 1986.

\bibitem{HawHunTay99}
S.~W. Hawking, C.~J. Hunter, and M.~M. Taylor-Robinson.
\newblock Rotation and the {AdS/CFT} correspondence.
\newblock {\em Phys. Rev. {\rm D}}, 59:064005, 1999.

\bibitem{Gibbonsetal04}
G.~W. Gibbons, H.~L{\"u}, D.~N. Page, and C.~N. Pope.
\newblock Rotating black holes in higher dimensions with a cosmological
  constant.
\newblock {\em Phys. Rev. Lett.}, 93:171102, 2004.

\bibitem{Pravdaetal04}
V.~Pravda, A.~Pravdov\'a, A.~Coley, and R.~Milson.
\newblock Bianchi identities in higher dimensions.
\newblock {\em Class. Quantum Grav.}, 21:2873--2897, 2004.
\newblock See also V. Pravda, A. Pravdov\'a, A. Coley and R. Milson {\em Class.
  Quantum Grav.} {\bf 24} (2007) 1691 (corrigendum).

\bibitem{ColPel06}
A.~Coley and N.~Pelavas.
\newblock Algebraic classification of higher dimensional spacetimes.
\newblock {\em Gen. Rel. Grav.}, 38:445--461, 2006.

\bibitem{Hamamotoetal07}
N.~Hamamoto, T.~Houri, T.~Oota, and Y.~Yasui.
\newblock Kerr-{NUT}-de~{S}itter curvature in all dimensions.
\newblock {\em J. Phys. A}, 40:F177--F184, 2007.

\bibitem{PraPraOrt07}
V.~Pravda, A.~Pravdov\'a, and M.~Ortaggio.
\newblock Type {D} {E}instein spacetimes in higher dimensions.
\newblock {\em Class. Quantum Grav.}, 24:4407--4428, 2007.

\bibitem{OrtPraPra09}
M.~Ortaggio, V.~Pravda, and A.~Pravdov\'a.
\newblock Higher dimensional {K}err-{S}child spacetimes.
\newblock {\em Class. Quantum Grav.}, 26:025008, 2009.

\bibitem{OrtPodZof08}
M.~Ortaggio, J.~Podolsk\'y, and M.~\v{Z}ofka.
\newblock {R}obinson-{T}rautman spacetimes with an electromagnetic field in
  higher dimensions.
\newblock {\em Class. Quantum Grav.}, 25:025006, 2008.

\bibitem{Coleyetal04}
A.~Coley, R.~Milson, V.~Pravda, and A.~Pravdov\'a.
\newblock Classification of the {W}eyl tensor in higher dimensions.
\newblock {\em Class. Quantum Grav.}, 21:L35--L41, 2004.

\bibitem{Coley08}
A.~Coley.
\newblock Classification of the {W}eyl tensor in higher dimensions and
  applications.
\newblock {\em Class. Quantum Grav.}, 25:033001, 2008.

\bibitem{OrtPraPra13rev}
M.~Ortaggio, V.~Pravda, and A.~Pravdov\'a.
\newblock Algebraic classification of higher dimensional spacetimes based on
  null alignment.
\newblock {\em Class. Quantum Grav.}, 30:013001, 2013.

\bibitem{EmpRea02prl}
R.~Emparan and H.~S. Reall.
\newblock A rotating black ring solution in five dimensions.
\newblock {\em Phys. Rev. Lett.}, 88:101101, 2002.

\bibitem{PraPra05}
V.~Pravda and A.~Pravdov\'a.
\newblock {WAND}s of the black ring.
\newblock {\em Gen. Rel. Grav.}, 37:1277--1287, 2005.

\bibitem{MonOCoWhi14}
R.~Monteiro, D.~O'Connell, and C.~D. White.
\newblock Black holes and the double copy.
\newblock {\em JHEP}, 12:056, 2014.

\bibitem{DerGur86}
T.~Dereli and M.~G{\"{u}}rses.
\newblock The generalized {K}err-{S}child transform in eleven-dimensional
  supergravity.
\newblock {\em Phys. Lett. {\rm B}}, 171:209--211, 1986.

\bibitem{ColHilSen01}
B.~Coll, S.~R. Hildebrandt, and J.~M.~M. Senovilla.
\newblock Kerr-{S}child symmetries.
\newblock {\em Gen. Rel. Grav.}, 33:649--670, 2001.

\bibitem{MalPra11}
T.~M\'alek and V.~Pravda.
\newblock {K}err-{S}child spacetimes with ({A})d{S} background.
\newblock {\em Class. Quantum Grav.}, 28:125011, 2011.

\bibitem{OrtSri24}
M.~Ortaggio and A~Srinivasan.
\newblock Charging {K}err-{S}child spacetimes in higher dimensions.
\newblock {\em Phys. Rev. {\rm D}}, 110:044035, 2024.

\bibitem{Srinivasan25}
A.~Srinivasan.
\newblock Algebraic and optical properties of generalized {K}err-{S}child
  spacetimes in arbitrary dimensions.
\newblock {\em Phys. Rev. {\rm D}}, 111:064061, 2025.

\bibitem{Durkee09}
M.~Durkee.
\newblock Type {II} {E}instein spacetimes in higher dimensions.
\newblock {\em Class. Quantum Grav.}, 26:195010, 2009.

\bibitem{DurRea09}
M.~Durkee and H.~S. Reall.
\newblock A higher-dimensional generalization of the geodesic part of the
  {G}oldberg-{S}achs theorem.
\newblock {\em Class. Quantum Grav.}, 26:245005, 2009.

\bibitem{ParWyl11}
A.~Garcia-Parrado G\'omez-Lobo and L.~Wylleman.
\newblock A new special class of {P}etrov type {D} vacuum space-times in
  dimension five.
\newblock {\em J. Phys.: Conf. Series}, 314:012024, 2011.

\bibitem{Ortaggioetal12}
M.~Ortaggio, V.~Pravda, A.~Pravdov\'a, and H.~S. Reall.
\newblock On a five-dimensional version of the {G}oldberg-{S}achs theorem.
\newblock {\em Class. Quantum Grav.}, 29:205002, 2012.

\bibitem{OrtPraPra13}
M.~Ortaggio, V.~Pravda, and A.~Pravdov\'a.
\newblock On the {G}oldberg-{S}achs theorem in higher dimensions in the
  non-twisting case.
\newblock {\em Class. Quantum Grav.}, 30:075016, 2013.

\bibitem{ReaGraTur13}
H.~S. Reall, A.~A.~H. Graham, and C.~P. Turner.
\newblock On algebraically special vacuum spacetimes in five dimensions.
\newblock {\em Class. Quantum Grav.}, 30:055004, 2013.

\bibitem{deFGodRea15}
G.~B. de~Freitas, M.~Godazgar, and H.~S. Reall.
\newblock Uniqueness of the {K}err-de~{S}itter spacetime as an algebraically
  special solution in five dimensions.
\newblock {\em Commun. Math. Phys.}, 340:291--323, 2015.

\bibitem{Wylleman15}
L.~Wylleman.
\newblock Finalizing the classification of type {II} or more special {E}instein
  spacetimes in five dimensions, arXiv:1511.02824 [gr-qc].

\bibitem{deFGodRea16}
G.~B. de~Freitas, M.~Godazgar, and H.~S. Reall.
\newblock Twisting algebraically special solutions in five dimensions.
\newblock {\em Class. Quantum Grav.}, 33:095002, 2016.

\bibitem{Ortaggio17}
M.~Ortaggio.
\newblock On the uniqueness of the {M}yers-{P}erry spacetime as a type {II(D)}
  solution in six dimensions.
\newblock {\em JHEP}, 06:042, 2017.

\bibitem{OrtPraPra18}
M.~Ortaggio, V.~Pravda, and A.~Pravdov\'a.
\newblock On higher dimensional {E}instein spacetimes with a non-degenerate
  double {W}eyl aligned null direction.
\newblock {\em Class. Quantum Grav.}, 35:075004, 2018.

\bibitem{TinPra19}
T.~Tint\v{e}ra and V.~Pravda.
\newblock On the {G}oldberg--{S}achs theorem in six dimensions.
\newblock {\em Gen. Rel. Grav.}, 51:111, 2019.

\bibitem{Taghavi-Chabert22}
A.~Taghavi-Chabert.
\newblock Twisting non-shearing congruences of null geodesics, almost {CR}
  structures and Einstein metrics in even dimensions.
\newblock {\em Ann. Mat. Pura Appl.}, 201:655--693, 2022.

\bibitem{MarPeo22}
M.~Mars and C.~Pe\'on-Nieto.
\newblock Covariant classification of conformal {K}illing vectors of locally
  conformally flat $n$-manifolds with an application to {K}err-de~{S}itter
  spacetimes.
\newblock {\em Phys. Rev. {\rm D}}, 106:084045, 2022.

\bibitem{OrtPraPra09b}
M.~Ortaggio, V.~Pravda, and A.~Pravdov\'a.
\newblock Asymptotically flat, algebraically special spacetimes in higher
  dimensions.
\newblock {\em Phys. Rev. {\rm D}}, 80:084041, 2009.

\bibitem{OrtPra14}
M.~Ortaggio and A.~Pravdov\'a.
\newblock Asymptotic behaviour of the {W}eyl tensor in higher dimensions.
\newblock {\em Phys. Rev. {\rm D}}, 90:104011, 2014.

\bibitem{PodOrt06}
J.~Podolsk\'y and M.~Ortaggio.
\newblock {R}obinson-{T}rautman spacetimes in higher dimensions.
\newblock {\em Class. Quantum Grav.}, 23:5785--5797, 2006.

\bibitem{CalEmpRod08}
M.~M. Caldarelli, R.~Emparan, and M.~J. Rodriguez.
\newblock Black rings in ({A}nti)-de~{S}itter space.
\newblock {\em JHEP}, 11:011, 2008.

\bibitem{ArmObe11}
J.~Armas and N.~A. Obers.
\newblock Blackfolds in (anti)-de~{S}itter backgrounds.
\newblock {\em Phys. Rev. {\rm D}}, 83:084039, 2011.

\bibitem{Caldarellietal12}
M.~M. Caldarelli, R.~G. Leigh, A.~C. Petkou, P.~M. Petropoulos, V.~Pozzoli, and
  K.~Siampos.
\newblock Vorticity in holographic fluids.
\newblock In {\em Proceedings of Proceedings of the Corfu Summer Institute 2011
  {\textemdash} PoS(CORFU2011)}, volume 155, page 076, 2012.

\bibitem{Gnecchietal14}
A.~Gnecchi, K.~Hristov, D.~Klemm, C.~Toldo, and O.~Vaughan.
\newblock Rotating black holes in 4d gauged supergravity.
\newblock {\em JHEP}, 01:127, 2014.

\bibitem{Klemm14}
D.~Klemm.
\newblock Four-dimensional black holes with unusual horizons.
\newblock {\em Phys. Rev. {\rm D}}, 89:084007, 2014.

\bibitem{HenManKub15}
R.~A. Hennigar, R.~B. Mann, and D.~Kubiz\v{n}\'ak.
\newblock Entropy inequality violations from ultraspinning black holes.
\newblock {\em Phys. Rev. Lett.}, 115:031101, 2015.

\bibitem{Hennigaretal15}
R.~A. Hennigar, D.~Kubiz\v{n}\'ak, R.~B. Mann, and N.~Musoke.
\newblock Ultraspinning limits and super-entropic black holes.
\newblock {\em JHEP}, 06:096, 2015.

\bibitem{MarPaeSen17}
M.~Mars, T.-T. Paetz, and J.~Senovilla.
\newblock Classification of {K}err--de~{S}itter-like spacetimes with
  conformally flat $\mathcal{I}$.
\newblock {\em Class. Quantum Grav.}, 34:095010, 2017.

\bibitem{MarPaeSen18}
M.~Mars, T.-T. Paetz, and J.~Senovilla.
\newblock Limit of {K}err--de~{S}itter spacetime with infinite angular-momentum
  parameter $a$.
\newblock {\em Phys. Rev. {\rm D}}, 97:024021, 2018.

\bibitem{MarPeo22_b}
M.~Mars and C.~Pe\'on-Nieto.
\newblock Classification of {K}err--de~{S}itter-like spacetimes with
  conformally flat $\mathcal{I}$ in all dimensions.
\newblock {\em Phys. Rev. {\rm D}}, 105:044027, 2022.

\bibitem{Ortaggio09}
M.~Ortaggio.
\newblock {B}el-{D}ebever criteria for the classification of the {W}eyl tensor
  in higher dimensions.
\newblock {\em Class. Quantum Grav.}, 26:195015, 2009.

\bibitem{OrtPraPra11}
M.~Ortaggio, V.~Pravda, and A.~Pravdov\'a.
\newblock On higher dimensional {E}instein spacetimes with a warped extra
  dimension.
\newblock {\em Class. Quantum Grav.}, 28:105006, 2011.

\bibitem{AshDas00}
A.~Ashtekar and S.~Das.
\newblock Asymptotically anti-de~{S}itter spacetimes: conserved quantities.
\newblock {\em Class. Quantum Grav.}, 17:L17--L30, 2000.

\bibitem{PraPra08}
A.~Pravdov\'a and V.~Pravda.
\newblock The {N}ewman-{P}enrose formalism in higher dimensions: vacuum
  spacetimes with a non-twisting geodetic multiple {W}eyl aligned null
  direction.
\newblock {\em Class. Quantum Grav.}, 25:235008, 2008.

\bibitem{GibWil87}
G.~W. Gibbons and D.~L. Wiltshire.
\newblock Space-time as a membrane in higher dimensions.
\newblock {\em Nucl. Phys. {\rm B}}, 287:717--742, 1987.

\bibitem{Birmingham99}
D.~Birmingham.
\newblock Topological black holes in anti-de {S}itter space.
\newblock {\em Class. Quantum Grav.}, 16:1197--1205, 1999.

\bibitem{CheLuPop06}
W.~Chen, H.~L{\"u}, and C.~N. Pope.
\newblock General {K}err-{NUT}-{AdS} metrics in all dimensions.
\newblock {\em Class. Quantum Grav.}, 23:5323--5340, 2006.

\bibitem{ChrConGra25}
P.~T. Chru\'{s}ciel, W.~Cong, and F.~Gray.
\newblock {K}err-{AdS} type higher dimensional black holes with non-spherical
  cross-sections of horizons.
\newblock {\em Class. Quantum Grav.}, 42:155007, 2025.

\bibitem{Coleyetal04vsi}
A.~Coley, R.~Milson, V.~Pravda, and A.~Pravdov\'a.
\newblock Vanishing scalar invariant spacetimes in higher dimensions.
\newblock {\em Class. Quantum Grav.}, 21:5519--5542, 2004.

\bibitem{OrtPraPra07}
M.~Ortaggio, V.~Pravda, and A.~Pravdov\'a.
\newblock Ricci identities in higher dimensions.
\newblock {\em Class. Quantum Grav.}, 24:1657--1664, 2007.

\bibitem{Durkeeetal10}
M.~Durkee, V.~Pravda, A.~Pravdov\'a, and H.~S. Reall.
\newblock Generalization of the {G}eroch-{H}eld-{P}enrose formalism to higher
  dimensions.
\newblock {\em Class. Quantum Grav.}, 27:215010, 2010.

\bibitem{Kubicek_thesis}
J.~Kub\'{\i}\v{c}ek.
\newblock {\em Gravitace ve vy\v{s}\v{s}\'{\i}ch dimenz\'{\i}ch}.
\newblock Diploma thesis, Charles University in Prague, 2015.

\bibitem{OrtPraPra10}
M.~Ortaggio, V.~Pravda, and A.~Pravdov\'a.
\newblock Type {III} and {N} {E}instein spacetimes in higher dimensions:
  general properties.
\newblock {\em Phys. Rev. {\rm D}}, 82:064043, 2010.

\bibitem{Gantmacherbook}
F.~R. Gantmacher.
\newblock {\em The theory of Matrices}, volume~1.
\newblock Chelsea Publishing Company, Providence, 1959.

\bibitem{Kokoska_PhD}
D.~Koko\v{s}ka.
\newblock {PhD} thesis, Charles University, Prague.
\newblock To appear.

\bibitem{BahLunWhi17}
N.~Bahjat-Abbas, A.~Luna, and C.~D. White.
\newblock The {K}err-{S}child double copy in curved spacetime.
\newblock {\em JHEP}, 12:004, 2017.

\bibitem{CarPenTro18}
M.~Carrillo-Gonz\'alez, R.~Penco, and M.~Trodden.
\newblock The classical double copy in maximally symmetric spacetimes.
\newblock {\em JHEP}, 04:028, 2018.

\bibitem{CheLu08}
W.~Chen and H.~L{\"u}.
\newblock {K}err-{S}child structure and harmonic 2-forms on
  ({A})ds-{K}err-{NUT} metrics.
\newblock {\em Phys. Lett. {\rm B}}, 658:158--163, 2008.

\bibitem{Krtous07}
P.~Krtou\v{s}.
\newblock Electromagnetic field in higher-dimensional black-hole spacetimes.
\newblock {\em Phys. Rev. {\rm D}}, 76:084035, 2007.

\bibitem{Aliev07}
A.N. Aliev.
\newblock Gyromagnetic ratio of charged {K}err--anti-de~{S}itter black holes.
\newblock {\em Class. Quantum Grav.}, 24:4669--4678, 2007.

\bibitem{Aliev07_2}
A.N. Aliev.
\newblock Electromagnetic properties of {K}err-anti-de~{S}itter black holes.
\newblock {\em Phys. Rev. {\rm D}}, 75:084041, 2007.

\bibitem{FroKrtKub17}
V.~P. Frolov, P.~Krtou\v{s}, and D.~Kubiz{\v n}{\'a}k.
\newblock Weakly charged generalized {K}err--{NUT}--{(A)dS} spacetimes.
\newblock {\em Phys. Lett. {\rm B}}, 771:254--256, 2017.

\bibitem{ChaKee23}
S.~Chawla and C.~Keeler.
\newblock Aligned fields double copy to {K}err-{NUT}-{(A)dS}.
\newblock {\em JHEP}, 04:005, 2023.

\bibitem{KolKrt17}
I.~Kol\'a\v{r} and P.~Krtou\v{s}.
\newblock {NUT}-like and near-horizon limits of {K}err-{NUT}-{(A)dS}
  spacetimes.
\newblock {\em Phys. Rev. {\rm D}}, 95:124044, 2017.

\bibitem{KleMorVan98}
D.~Klemm, V.~Moretti, and L.~Vanzo.
\newblock Rotating topological black holes.
\newblock {\em Phys. Rev. {\rm D}}, 57:6127--6137, 1998.
\newblock See also D. Klemm, V. Moretti, and L. Vanzo (1999), Erratum: Rotating
  topological black holes [Phys. Rev. D 57, 6127 (1998)], {\em Phys. Rev.} D
  60:109902.

\bibitem{Klemm98}
D.~Klemm.
\newblock Rotating black branes wrapped on {E}instein spaces.
\newblock {\em JHEP}, 11:019, 1998.

\bibitem{Ortaggio07}
M.~Ortaggio.
\newblock Higher dimensional spacetimes with a geodesic, shearfree, twistfree
  and expanding null congruence.
\newblock In {\em Proceedings of the XVII {SIGRAV} Conference (Torino,
  September 4--7, 2006)}, 2007. gr-qc/0701036.

\bibitem{Kerr63}
R.~P. Kerr.
\newblock Gravitational field of a spinning mass as an example of algebraically
  special metrics.
\newblock {\em Phys. Rev. Lett.}, 11:237--238, 1963.

\bibitem{MarPeo21}
M.~Mars and C.~Pe\'on-Nieto.
\newblock Free data at spacelike $\mathcal{I}$ and characterization of
  {K}err-de~{S}itter in all dimensions.
\newblock {\em Eur. Phys. J.~C}, 81:914, 2021.

\end{thebibliography}

\end{document}